\documentclass[numberedappendix]{emulateapj}

\usepackage{multirow}
\usepackage{graphicx}
\usepackage{natbib}
\usepackage{amsmath}

\begin{document}

\title{Close Stellar Encounters in Young, Substructured, Dissolving Star Clusters: Statistics and Effects on Planetary Systems}

\author{Jonathan Craig \& Mark R.~Krumholz}
\affil{Department of Astronomy \& Astrophysics, University of California, Santa 
Cruz, CA 95064 USA}
\email{krumholz@ucolick.org}
\begin{abstract}
Both simulations and observations indicate that stars form in filamentary, hierarchically clustered associations, most of which disperse into their galactic field once feedback destroys their parent clouds. However, during their early evolution in these substructured environments, stars can undergo close encounters with one another that might have significant impacts on their protoplanetary disks or young planetary systems. We perform N-body simulations of the early evolution of dissolving, substructured clusters with a wide range of properties, with the aim of quantifying the expected number and orbital element distributions of encounters as a function of cluster properties. We show that the presence of substructure both boosts the encounter rate and modifies the distribution of encounter velocities compared to what would be expected for a dynamically relaxed cluster. However, the boost only lasts for a dynamical time, and as a result the overall number of encounters expected remains low enough that gravitational stripping is unlikely to be a significant effect for the vast majority of star-forming environments in the Galaxy. We briefly discuss the implications of this result for models of the origin of the Solar System, and of free-floating planets. We also provide tabulated encounter rates and orbital element distributions suitable for inclusion in population synthesis models of planet formation in a clustered environment.
\end{abstract}


\section{Introduction}


Stars form in giant molecular clouds (GMCs) that possess a high degree of substructure. They tend to be clumpy and filamentary \citep{1994ApJ...428..693W,2000prpl.conf...97W}, almost certainly as a result of pervasive supersonic turbulence \citep{1981MNRAS.194..809L,2004RvMP...76..125M}. Stars that form out of these clouds inherit the substructures of their parents, leading to a hierarchy of clustering \citep{2003ARA&A..41...57L, bressert10a, gutermuth11a}, and to self-similar (fractal) structures within star clusters \citep{1995MNRAS.272..213L,2000ApJ...530..277E,2001AJ....121.1507E,2004MNRAS.348..589C,2005ApJ...619..779C}. Numerical simulations of star formation produce similar results \citep{2000ApJS..128..287K,2003MNRAS.343..413B, 2009ApJ...704L.124O, krumholz12b}.

\citet{1972A&A....21..255A} were the first to study the evolution of star clusters with initial substructure. They found that any substructure initially present was typically destroyed within one free-fall time, and observations generally support this picture \citep{2004MNRAS.348..589C,2008MNRAS.389.1209S}. However, simulations indicate that substructure can be destroyed quickly only in initially subvirial clusters, a case characterized by an initial collapse of the star cluster towards the center of mass followed by a chaotic evolutionary phase \citep{2004A&A...413..929G,2010MNRAS.407.1098A}. For supervirial clusters, on the other hand, substructure can survive for up to several crossing times \citep{2004A&A...413..929G}.

Both observations and theory suggest that clusters typically form subvirially with respect to the gas, though not necessarily with respect to the stars alone \citep{2008ApJ...676.1109F,tobin09a, 2009ApJ...704L.124O}. However, the star formation process is inefficient, with relatively small amounts of the mass in a given molecular cloud being converted into stars, which then expel the remainder of the cloud into back into the diffuse interstellar medium through their radiation, winds, and supernovae \citep{1980ApJ...235..986H, LADA1999,2003ARA&A..41...57L, matzner02a, krumholz06d, 2010ApJ...710L.142F, goldbaum11a, kruijssen12a}. Once the gas is expelled, stars disperse into the field, with only a minority remaining in bound clusters for many dynamical times after gas dispersal. As a result, even if stars are born subvirial with respect to the gas, they may be rendered supervirial by its rapid dispersal. Real star clusters may therefore experience periods of both subvirial and supervirial evolution.\footnote{A brief comment on terminology: some authors use the word ``cluster" to refer to any significant stellar over density regardless of its dynamical state, while others use the term to refer exclusively to stellar structures that remain bound after gas dispersal. We follow the former approach, and refer to our objects as clusters even though they are unbound.}

Whether subvirial or supervirial, this early evolutionary stage is of considerable interest for the problem of planet formation. In denser environments and massive clusters containing massive stars, where a significant fraction of stars appear to form \citep{2003ARA&A..41...57L, chandar10a}, close passages between Solar-type stars and massive stars may lead to the photoevaporation of protoplanetary disks and modification of the planet formation process \citep[and references therein]{2004ApJ...611..360A, throop05a, 2010ARA&A..48...47A}. Close encounters with passing stars can also gravitationally disrupt both disks and planetary systems, potentially truncating disks, exciting planetary orbits, or ejecting planets completely. Such encounters have been suggested as a potential explanation for such diverse observations as the existence of free-floating planets \citep[e.g.][]{2011Natur.473..349S, 2012MNRAS.421L.117V} and the structure of the Kuiper Belt \citep[e.g.][]{lestrade11a, jimenez-torres11a}. The need to avoid disruptive encounters has also been used as a constraint on the potential birth environment of the Sun \citep{2001Icar..150..151A, 2006ApJ...641..504A}. Our Solar System is remarkably well ordered compared to many extrasolar planetary systems, with all of the planets on nearly circular orbits (every planet except Mercury has $e<0.09$), while the Kuiper belt is also relatively undisturbed. \citet{2001Icar..150..151A} and \citet{2010ARA&A..48...47A} have argued that this implies that the Sun could have been formed in a cluster no larger than $\sim 10^3$ stars, though this conclusion has recently been questioned by \citet{Dukes}.

A crucial input to all these questions is the rate and distribution of orbital elements of the encounters that a star in a cluster will experience. These are a necessary ingredient for population synthesis models for planet formation in clustered environments \citep[e.g.][]{2010ARA&A..48...47A,Dukes, 2012arXiv1209.2136O}. While a number of authors have measured these distributions numerically \citep[e.g.][]{2001MNRAS.322..859B,2006ApJ...641..504A,2009ApJ...697..458S,2006ApJ...642.1140O,2010A&A...509A..63O}, none thus far have done so in the context of a dispersing, initially highly-substructured cluster, which modern observations suggest is the typical condition for the formation of most stars. 

Several authors have recently studied properties of initially substructured 
(fractal) clusters in slightly different contexts. \citet{2011MNRAS.418.2565P} find that reproducing the observed present-day binary properties of the ONC  require that it have formed with a high degree of substructure and a high initial binary fraction. \citet{2012MNRAS.427..637P} find that the surface density of fractal star clusters decreases rapidly in time, which implies that a large fraction of star-star encounters will occur very early on in the cluster life. \citet{2013MNRAS.428.1303S} found that gas removal after multiple crossing times typically results in relaxed clusters, with no remaining substructure, whereas quick gas expulsion before one crossing time leads to a stochastic, unpredictable outcome. 

The two papers closest to our work are \citet{2006ApJ...641..504A}, who calculate encounter rates in both dispersing and cold clusters, but do not include any initial substructure, and \citet{2012MNRAS.419.2448P} who do include substructure and study how star clusters affects orbital elements of planetary systems. We add to these studies by conducting a series of N-body simulations of dispersing, fractal star clusters across a much broader parameter space than has been considered before. We consider a wide range of dynamical environments, from unbound, supervirial stellar associations to subvirial stars in a gas-dominated clump that are subsequently unbound by gas expulsion. For each simulation we track every event in which two or three stars pass within $1000$ AU of one another, giving a nearly complete dynamic profile of possible interactions. Our work expands on that of \citet{2006ApJ...641..504A} and \citet{2012MNRAS.419.2448P} by surveying a significantly broader parameter space, with cluster masses from $30 - 30,000$ $M_\odot$ and surface densities from $0.1 - 3.0$ g cm$^{-2}$. This broad survey allows us to measure how the results depend on cluster mass and surface density, and thereby to extrapolate into the regime of high mass and surface density clusters that are too computationally expensive to simulate directly.

The remainder of this paper is as follows. Section \ref{sec:methods} discusses the model parameters, the initial conditions for the clusters, and the simulations and data reduction methods, and defines the statistical distributions of interest. Section \ref{sec:results} details our results, and Section \ref{sec:discussion} discusses their implications. Our conclusions are presented in section \ref{sec:conclusion}. 

\section{Methods}
\label{sec:methods}

To study stellar encounters in dissolving clusters, we perform an ensemble of N-body simulations using a modified version of the numerical integrator NBODY6 \citep{1999PASP..111.1333A}. Below we describe the parameters in our simulations, the initial conditions, and how we process the resulting data.

\subsection{Simulation Parameters and Initial Conditions}

We characterize clusters by four parameters: the virial ratio, $Q$, defined as the ratio of kinetic to potential energy (so that a cluster in equilibrium has $Q=0.5$), the fractal dimension $D$, the stellar mass $M_c$, and the cluster surface density, $\Sigma_c$. We describe below how we use these parameters to set up the initial conditions in our simulations. We consider four combinations of $Q$ and $D$, and for each combination we then simulate clusters with a broad range of masses $M_c$ and surface densities $\Sigma_c$.


We use the surface density $\Sigma_c$ rather than the radius $R_c$ or the volume density $\rho_c$ as a parameter for two reasons. First, while the volume density determines encounter rates, the quantity of interest for the standpoint of studying how clusters affect planetary systems is the total number of encounters a star can expect to experience over the cluster lifetime, not the encounter rate. The natural time scale for a disrupting cluster is the crossing time, and the total number of encounters per crossing time depends on the surface density rather than the volume density \citep{Dukes}. Second, observations of cluster-forming gas clumps appear to indicate that, while clusters span a very wide range of volume densities and radii, they form a sequence of relatively constant surface density \citep[and references therein]{2010ApJ...710L.142F}. Thus in discussing embedded clusters that have until recently been dominated by the potential of the gas, it is also natural to work in terms of surface rather than volume density.

Our base case is a cluster with $Q=0.75$ and $D=2.2$; this value of virial ratio corresponds approximately to a cluster that has just expelled its residual gas but has not been completely unbound by the process, and the fractal dimension describes a cluster with a moderate degree of substructure. This is consistent with observations which have typically found that $D$ goes from 1.9 to 2.5 \citep{1991ApJ...378..186F,1994AA...291..557V,1996ApJ...471..816E,
2006A&A...452..163D,2010ApJ...720..541S}. These runs generally result in majority of the stars escaping promptly, but some remaining as a bound structure for long times. Our second case is $Q=1.25$ and $D=2.2$, corresponding to a cluster that has been completely unbound by gas expulsion, or that was never bound in the first place. In these runs essentially all the stars disperse. Our third case is a model with $D=1.6$ with $Q=0.75$, corresponding to a case like the base model but with more substructure. In our fourth model, we explicitly include a phase in which the stars are confined by an external potential, which we rapidly remove after four crossing times, where the crossing time is
\begin{equation}
t_c = \frac{M_c^{1/4}}{G^{1/2} (\pi \Sigma_c)^{3/4}}.
\end{equation}
Our motivation for this choice is that observations indicate that the lifetime of the embedded phase of star cluster formation is roughly four crossing times \citep{tan06a}, or possibly even less \citep{2000ApJ...530..277E}. We should note here that substructure is typically erased after several crossing times in a confined potential  \citep{2013MNRAS.428.1303S}. The first three cases assume that the gas is expelled very early while the substructure still remains. While it is present, we describe the gas potential by a Plummer model,\begin{equation}
\Phi(r) = - \frac{GM_{\rm gas}}{r_c}\frac{1}{\sqrt{1+(\frac{r}{r_c})^2}},
\end{equation}
and we choose $M_{\rm gas} = (0.7/0.3) M_c$, so that the gas mass is $70\%$ of the total gas plus stellar mass. 
This cluster has $Q = 0.3$ and $D=2.2$ initially. We should note that this value of the virial ratio is computed using only the potential energy due to the interactions between stars, not the coupling of the stars to the gas. The total potential energy is
\begin{equation}
U_{\rm tot} = U_{*,*} + U_{*,g} = -\sum_{i=1}^N \sum_{j=i+1}^N \frac{G m_i m_j}{r_{ij}} - \sum_{i=1}^N m_i \Phi(r_i)
\end{equation}
where $N$ is the number of stars, $r_{ij}$ is the distance between the $i^{\rm th}$ and $j^{\rm th}$ stars, and $r_i$ is the radial position of the $i^{\rm th}$ star. This means that the real virial ratio is then $T / U_{\rm tot}$. The gas mass term dominates, which leaves us with $Q < 0.1$, an extremely subvirial case. This effectively gives us a bound on how important the effect of gas might be on the evolution of the cluster.


We summarize the model parameters in Table \ref{tab:models}, and the number of independent realizations we perform at each $(M_c,\Sigma_c)$ combination in Tables \ref{NRUND2.2Q0.75} -- \ref{NRUND2.2Q0.3}. The numbers of runs for each $(M_c,\Sigma_c)$ value are chosen so that the number of interactions, and thus the statistical error on our results, is roughly constant. This implies a large number of runs for small, low surface density cases, and a smaller number of runs for more massive, higher surface density cases. As we will see below when we discuss our error budget, this does limit our accuracy to some extent in the high mass and surface density regime. Unfortunately this regime is too computationally-costly to allow a significantly larger number of simulations. The relatively small number of simulations limits our accuracy, but even with this limitation we show below that the errors on our measured encounter rates are typically no more than $\sim 10\%$.


\begin{deluxetable}{cccc}
\tablecaption{
\label{tab:models}
Model parameters
}
\tablehead{
\colhead{Name} &
\colhead{$Q$} &
\colhead{$D$} &
\colhead{Gas?}
}
\startdata
Q0.75D2.2 & 0.75 & 2.2 & No \\
Q1.25D2.2 & 1.25 & 2.2 & No \\
Q0.75D1.6 & 0.75 & 1.6 & No \\
Gas & 0.3 & 2.2 & Yes
\enddata
\end{deluxetable}

\begin{deluxetable}{cc}
\tablecaption{
\label{tab:MN}
Cluster mass and number of stars.
}
\tablehead{
\colhead{log$(M_c / M_{\odot})$} &
\colhead{$N=M_c/\bar{m}$} 
}
\startdata
 1.5 & 54\\
 2.0 & 170 \\
 2.5 & 540 \\
 3.0 & 1707 \\
 3.5 & 5400 \\
 4.0 & 17077 \\
 4.5 & 54003
\enddata
\end{deluxetable}

\begin{deluxetable}{crrrrrrr}
\tablecaption{
\label{NRUND2.2Q0.75}
Number of realizations for model Q0.75D2.2
}
\tablehead{
\colhead{$\Sigma_c$} &
\multicolumn{7}{c}{$\log (M_c/M_\odot)$ } \\
\colhead{[g cm$^{-2}$]} &
\colhead{1.5} &
\colhead{2.0} &
\colhead{2.5} &
\colhead{3.0} &
\colhead{3.5} &
\colhead{4.0} &
\colhead{4.5}
}
\startdata
3.0 & 600 & 150 & 40 & 8 & 5 & 2 & 0 \\
1.0 & 1000 & 200 & 50 & 15 & 5 & 4 & 3  \\
0.5 & 1500 & 300 & 75 & 20 & 10 & 5 & 3  \\
0.1 & 4000 & 800 & 150 & 40 & 20 & 5 & 4
\enddata
\end{deluxetable}

\begin{deluxetable}{crrrrr}
\tablecaption{
\label{NRUND2.2Q1.25}
Number of realizations for model Q1.25D2.2
}
\tablehead{
\colhead{$\Sigma_c$} &
\multicolumn{5}{c}{$\log (M_c/M_\odot)$ } \\
\colhead{[g cm$^{-2}$]} &
\colhead{1.5} &
\colhead{2.0} &
\colhead{2.5} &
\colhead{3.0} & 
\colhead{3.5}
}
\startdata
3.0 & 300 & 100 & 20 & 10 & 3\\
1.0 & 500 & 175 & 30 & 15 & 4 \\
0.5 & 700 & 175 & 40 & 15 & 4 \\
0.1 & 1000 & 275 & 75 & 30 & 10 
\enddata
\end{deluxetable}

\begin{deluxetable}{crrrrr}
\tablecaption{
\label{NRUND1.6Q0.75}
Number of realizations for model Q0.75D1.6
}
\tablehead{
\colhead{$\Sigma_c$} &
\multicolumn{5}{c}{$\log (M_c/M_\odot)$ } \\
\colhead{[g cm$^{-2}$]} &
\colhead{1.5} &
\colhead{2.0} &
\colhead{2.5} &
\colhead{3.0} & 
\colhead{3.5}
}
\startdata
1.0 & 500 & 100 & 15 & 5 & 0 \\
0.5 & 500 & 100 & 30 & 6 & 0 \\
0.1 & 500 & 150 & 40 & 8 & 5
\enddata
\end{deluxetable}

\begin{deluxetable}{crrrrrrr}
\tablecaption{
\label{NRUND2.2Q0.3}
Number of realizations for model Gas
}
\tablehead{
\colhead{$\Sigma_c$} &
\multicolumn{5}{c}{$\log (M_c/M_\odot)$ } \\
\colhead{[g cm$^{-2}$]} &
\colhead{1.5} &
\colhead{2.0} &
\colhead{2.5} &
\colhead{3.0} & 
\colhead{3.5} &
\colhead{4.0} 
}
\startdata
3.0 & 175 & 50 & 15 & 6 & 3 & 0\\ 
1.0 & 225 & 75 & 25 & 8 & 4 & 2\\
0.5 & 275 & 100 & 30 & 10 & 5 & 3\\
0.1 & 350 & 125 & 40 & 12 & 6 & 4
\enddata
\end{deluxetable}

\subsection{Initial Conditions}



We initialize our clusters using the fractal initial conditions model with slight modifications \citep{2002MNRAS.334..156S,2004A&A...413..929G}. We refer the reader to the second paper for full details of the method, which we briefly summarize below. To generate the cluster we start by defining a cube with sides of length 2 (in arbitrary units, which will be scaled later to give the correct physical units), centered at the origin. This cube is subdivided into the 8 Cartesian sectors, and a first generation particle is placed at the center of each subcube. Each of these first generation cubes is then subdivided again, with second generation particles placed at the center of each second generation subcube. We repeat this subdivision procedure, with one additional constraint for the second and subsequent generations: each parent particle only has a probability $2^{D-3}$ of producing offspring. When $D=3$ this ensures that all positions are equally populated and there is no substructure, but for $D<3$ parts of the cube will be empty, yielding substructure. At each generation $g$, we also add a random displacement of position of magnitude $2^{-g-1}$ to prevent the development of an overly gridded structure. 

We repeat this procedure until the number of particles generated greatly exceeds the number we will actually use in the simulation. We then randomly select a subset of the points with radius less than 1 (in our arbitrary units) to be the initial locations of our stars. The radial positions of the stars are then 
multiplied by a factor
\begin{equation}
r_c= \sqrt{\frac{M_c}{\pi \Sigma_c}}.
\end{equation}
so that the average surface density of the cluster is $\Sigma_c$.

The number of stars is simply $M_c/\bar{m}$, where $\bar{m}$ is the mean stellar mass for our chosen IMF (see below). Table~\ref{tab:MN} gives the correspondence between the cluster mass and the number of stars $N$. Note that this means that for a given $M_c$ the actual cluster mass may be slightly larger or smaller, depending on drawing from the IMF. We therefore interpret $M_c$ as the expectation value of the cluster mass, though deviations from this value are small as long as $M_c \gg \bar{m}$.

We assign initial velocities to the stars using a recursive procedure to ensure that positions and velocities are correlated, as suggested by observations and simulations. At each generation we assign a random scalar velocity drawn from a Maxwellian distribution to each particle
\begin{equation}
\label{Maxwell}
p(v,g) \propto v^2 \exp\left(-2^{2g}\frac{v^2}{2\sigma_{v,0}^2}\right) .
\end{equation}
The direction of the velocity vector is chosen randomly. A particle's velocity is the value produced by this drawing added to the velocity of its parent.  Since the magnitude of the velocity perturbation decays with generation, the positions of the stars are then highly dependent on the velocities of the first few parents. Note that the choice of $\sigma_{v,0}^2$ is arbitrary, since we scale the final speeds so that the cluster has a specified virial ratio (see below). Figure \ref{fig:coherent} shows an example of a cluster generated via this procedure. 

\begin{figure}
\plotone{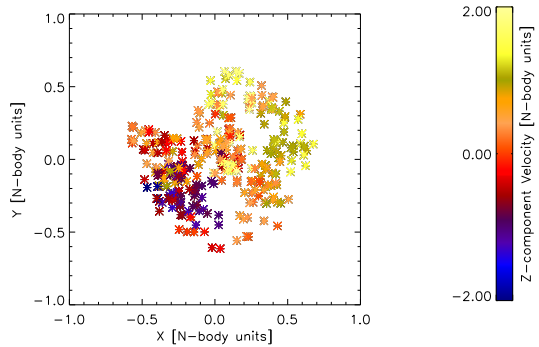}
\caption{
Asterisks indicate the positions of stars in an example cluster projected onto the $xy$-plane; colors indicate stars' $z$ velocities. Notice that velocities and positions are correlated.}
\label{fig:coherent}

\end{figure}

Finally, we assign stellar masses by randomly drawing from an extended version of the \citet{2002Sci...295...82K} IMF,
\begin{equation}
p(m) \propto 
 \begin{cases}
 m^{-0.3}, & 0.08 \le m/M_{\odot} < 0.1 \\
 m^{-1.3}, & 0.1 \le m/M_{\odot} < 0.5 \\ 
 m^{-2.3}, & 0.5 \le m/M_{\odot} \le 120
 \end{cases}
 .
\end{equation}
This IMF yields a mean stellar mass $\bar{m} \approx 0.59$ $M_{\odot}$. Once we have drawn all the stellar masses, positions, and velocities, we scale the velocities by a constant factor so that the initial virial ratio is the desired value.

\subsection{Simulations and Analysis}


We simulate the evolution of each cluster for 5 crossing times, except for the Gas runs, which we compute for 9 crossing times (i.e.~5 crossing times after the gas potential is removed). Since we are interested in close encounters for Solar-like stars, we track every instance in which a star of mass $0.8-1.2$ $M_\odot$ passes within $1000$ AU of another star and the pair has a non-negative center of mass energy; the latter condition excludes cases where the two stars form a binary. In addition, once we have found a pair of stars that meet this criterion, we also check for any other star passing within $1000$ AU of either body. 

The raw data we obtain from NBODY6 is a list of 2-body and 3-body interactions with positions, masses, velocities, indices and the time. Almost all interactions are recorded as a time series, since the stars involved are within 1000 AU of one another for more than a single simulation time step. Our end goal is to use these time series to calculate the distribution of impact parameters $b$ and relative velocities at infinity $v_{\infty}$ for encounters in clusters. For a given pair of particle positions and velocities, it is straightforward to compute what values of $b$ and $v_{\infty}$ would be required to produce that particular separation and relative velocity. However, in practice interactions are often complex, particularly in the high surface density cases where stars are tightly packed, and the values of $b$ and $v_\infty$ that one computes in this manner are not constant over the time series of positions and velocities that describes a particular interaction. For this reason, we must first classify interactions in order to decide how to analyze them. Our classification scheme is as follows.

\paragraph{1+1 interactions} These are true single star-by-single star scattering events. Two bodies are said to be well-described as a 1+1 interaction if, for that pair of particles, the set of impact parameters $b_k$ and relative velocities $v_{\infty,k}$ we compute from our time series satisfy the condition that 
\begin{equation}
\frac{\sigma_b}{\bar{b}} < T_{1+1} , \qquad \frac{\sigma_{v_{\infty}}}{\bar{v}_{\infty}} < T_{1+1},
\end{equation} 
where $\sigma_b$ and $\bar{b}$ are the standard deviation and mean of the time series $b_k$ and similarly for $\sigma_{v_{\infty}}$ and $\bar{v}_{\infty}$, and $T_{1+1} = 0.1$ is the tolerance ratio we adopt for 1+1 interactions. This value is somewhat arbitrary, but provides a reasonable separation between cases where two interacting stars have their orbits perturbed slightly by the potential of other stars during the interaction, and cases where another nearby star provides a large perturbation to the orbits. For 1+1 interactions, we record $\bar{b}$ and $\bar{v}_{\infty}$ as the impact parameter and relative velocity of the encounter.

\paragraph{1+2 interactions} These are events in which a single star scatters off a binary. We label an encounter involving three stars within 1000 AU of one another as a 1+2 interaction if two conditions are met. First, exactly one pair of the three stars must be gravitationally bound, while the other pairs are unbound. Second, if we replace this gravitationally bound pair by a single star at the center of mass position and velocity of the pair, and with a mass equal to the sum of the pair's masses, and we compute a set of impact parameters $b_k$ and relative velocities $v_{\infty,k}$ between this binary and the remaining star, we find that 
\begin{equation}
\frac{\sigma_b}{\bar{b}} < T_{1+2} , \qquad \frac{\sigma_{v_{\infty}}}{\bar{v}_{\infty}} < T_{1+2},
\end{equation} 
where $T_{1+2} = 0.3$. We set the tolerance somewhat higher than in the 1+1 case because, even in the absence of perturbations from external stars, tidal forces exerted by the binary on the single star may lead to some exchange of energy and angular momentum between the binary's internal energy and angular momentum and that of the orbit of the binary and the single star about one another. For 1+2 interactions, we record $\bar{b}$ and $\bar{v}_{\infty}$ as the impact parameter and relative velocity of the encounter.

\paragraph{1+1+1 interactions} These are events in which three unbound stars encounter one another. We classify an event as 1+1+1 if no pair composed of two of the three stars involved is mutually gravitationally bound. We decompose encounters of this type into three 1+1 events; since these 1+1 events clearly will not satisfy the tolerance criteria for 1+1 events, we simply calculate $b$ and $v_\infty$ in these cases using the positions and velocities the stellar pairs have at their point of closest approach.

\paragraph{Complex interactions} These are events which do not fall into one of the above categories. They may, for example, be cases where a metastable hierarchical multiple star systems forms and then dissolves some time later. These interactions do not have well-defined orbital elements. We do not attempt to define an impact parameter or relative velocity in these cases, and we do not include them in our statistical distributions of $b$ and $v_\infty$. We do, however, record such interactions and include them in our total counts of events.

\subsection{Statistical Distributions}

For each set of simulations, we are interested in three quantities. The first is simply the expected number of encounters $N_{\rm enc}$ within 1000 AU. The other two are the distributions of impact parameters $p(b)$ and relative velocities $p(v_\infty)$ that describe these encounters. For a fully relaxed cluster, we expect these to follow
\begin{equation}
p(b) \propto b,  
\label{bdistri}
\end{equation}
and
\begin{equation}
p(v_{\infty}) \propto v_{\infty}^2 \exp \left(-\frac{v^2_{\infty}}{2 \sigma_v^2}\right),
\end{equation}
but they need not for fractal, dispersing clusters that have not had time to relax. To evaluate the distributions from our simulations, we bin all our encounters into $N_{\rm bin} = 20$ equally-spaced bins of impact parameter from $0-1000$ AU, and into $N_{\rm bin}$ equally-spaced bins of relative velocity from $0-20$ km s$^{-1}$.

\section{Results}
\label{sec:results}

\begin{deluxetable*}{ccrrrrr}
\tablecaption{
\label{EnstatBASE}
Encounter Statistics for model Q0.75D2.2}
\tablehead{
\colhead{$\log(M_c/M_{\odot})$} &
\colhead{$\Sigma_c$} &
\colhead{$N_{\rm enc}$} &
\colhead{$\sigma_{\rm Poisson}$} &
\colhead{$\sigma_{\rm sample}$} &
\colhead{$v_{\infty}^{\rm median}$} &
\colhead{$\sigma_r$} \\
\colhead{} &
\colhead{[g cm$^{-2}$]} &
\colhead{} &
\colhead{} &
\colhead{} &
\colhead{ [km s$^{-1}$]}&
\colhead{}
}
\startdata
 1.5 & 0.1 & 0.85 & 0.01 & 0.00 & 1.88 & 0.01 \\
 1.5 & 0.5 & 3.82 & 0.03 & 0.00 & 2.71 & 0.01\\
 1.5 & 1.0 & 7.08 & 0.04 & 0.01 & 3.20 & 0.01\\
 1.5 & 3.0 & 9.64 & 0.07 & 0.01 & 4.23 & 0.01 \\
 2.0 & 0.1 & 1.78 & 0.01 & 0.00 & 2.48 & 0.01 \\
 2.0 & 0.5 & 8.25 & 0.05 & 0.02 & 3.62 & 0.01 \\
 2.0 & 1.0 & 13.69 & 0.08 & 0.04 & 4.18 & 0.01 \\
 2.0 & 3.0 & 30.21 & 0.14 & 0.10 & 5.25 & 0.01 \\
 2.5 & 0.1 & 2.87 & 0.02 & 0.01 & 3.06 & 0.01 \\
 2.5 & 0.5 & 13.63 & 0.07 & 0.08 & 4.60 & 0.01 \\
 2.5 & 1.0 & 23.83 & 0.12 & 0.20 & 5.13 & 0.01 \\
 2.5 & 3.0 & 76.39 & 0.23 & 0.89 & 6.79 & 0.01 \\
 3.0 & 0.1 & 3.46 & 0.03 & 0.04 & 3.41 & 0.01\\
 3.0 & 0.5 & 17.92 & 0.09 & 0.54 & 5.34 & 0.03 \\
 3.0 & 1.0 & 34.59 & 0.15 & 0.77 & 6.25 & 0.02 \\
 3.0 & 3.0 & 129.41 & 0.38 & 10.25 & 7.37 & 0.08 \\
 3.5 & 0.1 & 4.82 & 0.03 & 0.06 & 4.02 & 0.01\\
 3.5 & 0.5 & 31.00 & 0.10 & 1.22 & 6.65 & 0.04 \\
 3.5 & 1.0 & 77.44 & 0.22 & 4.58 & 7.58 & 0.06\\
 3.5 & 3.0 & 160.27 & 0.31 & 6.49 & 9.92 & 0.04 \\
 4.0 & 0.1 & 5.72 & 0.03 & 0.42 & 4.35 & 0.07\\
 4.0 & 0.5 & 47.44 & 0.09 & 3.88 & 7.72 & 0.08 \\
 4.0 & 1.0 & 92.48 & 0.15 & 8.95 & 9.09 & 0.10\\
 4.0 & 3.0 & 367.63 & 0.43 & 60.44 & 11.59 & 0.16\\
 4.5 & 0.1 & 9.01 & 0.03 & 0.10 & 5.47 & 0.01 \\
 4.5 & 0.5 & 46.39 & 0.07 & 3.65 & 8.29 & 0.08 \\
 4.5 & 1.0 & 71.23 & 0.09 & 10.13 & 8.87 & 0.14
 \enddata
 \tablecomments{$N_{\rm enc}$ is the mean number of encounters within $1000$ AU per star for a Sun-like star over the full duration of the simulation. $\sigma_{\rm Poisson}$ and $\sigma_{\rm sample}$ are the Poisson and parameter space sampling errors on $N_{\rm enc}$, and $\sigma_r$ is the total relative error $\delta N_{\rm enc}/N_{\rm enc}$ considering both sources; see Equations (\ref{eq:sigmapoisson}), (\ref{eq:sigmasample}), and (\ref{eq:sigmar}). $v_{\infty}^{\rm median}$ is the median encounter velocity.
 }
\end{deluxetable*}

\begin{deluxetable*}{crrrrrr}
\tablecaption{
\label{EnstatSUPER}
Encounter Statistics for model Q1.25D2.2}
\tablehead{
\colhead{$\log(M_c/M_{\odot})$} &
\colhead{$\Sigma_c$} &
\colhead{$N_{\rm enc}$} &
\colhead{$\sigma_{\rm Poisson}$} &
\colhead{$\sigma_{\rm sample}$} &
\colhead{$v_{\infty}^{\rm median}$} &
\colhead{$\sigma_r $} \\
\colhead{} &
\colhead{[g cm$^{-2}$]} &
\colhead{} &
\colhead{} &
\colhead{} &
\colhead{ [km s$^{-1}$]} &
\colhead{}
}
\startdata
 1.5 & 0.1 & 0.47 & 0.01 & 0.00 & 1.85 & 0.02 \\
 1.5 & 0.5 & 2.04 & 0.03 & 0.00 & 2.73 & 0.01 \\
 1.5 & 1.0 & 3.55 & 0.04 & 0.01 & 3.17 & 0.01 \\
 1.5 & 3.0 & 9.93 & 0.10 & 0.03 & 4.30 & 0.01\\
 2.0 & 0.1 & 1.08 & 0.02 & 0.00 & 2.31 & 0.02\\
 2.0 & 0.5 & 5.34 & 0.05 & 0.02 & 3.51 & 0.01\\
 2.0 & 1.0 & 8.50 & 0.07 & 0.03 & 4.17 & 0.01\\
 2.0 & 3.0 & 21.40 & 0.13 & 0.10 & 5.19 & 0.01\\
 2.5 & 0.1 & 2.27 & 0.03 & 0.02 & 2.93 & 0.02 \\
 2.5 & 0.5 & 8.31 & 0.08 & 0.09 & 3.65 & 0.01 \\
 2.5 & 1.0 & 14.05 & 0.12 & 0.38 & 4.80 & 0.03 \\
 2.5 & 3.0 & 63.96 & 0.30 & 1.44 & 6.52 & 0.02 \\
 3.0 & 0.1 & 2.16 & 0.03 & 0.03 & 3.10 & 0.02 \\
 3.0 & 0.5 & 9.31 & 0.08 & 0.19 & 4.65 & 0.02\\
 3.0 & 1.0 & 28.44 & 0.13 & 1.35 & 5.53 & 0.05 \\
 3.0 & 3.0 & 68.52 & 0.25 & 3.64 & 7.16 & 0.05\\
 3.5 & 0.1 & 3.30 & 0.03 & 0.08 & 3.70 & 0.03\\
 3.5 & 0.5 & 22.07 & 0.13 & 2.62 & 5.79 & 0.12\\
 3.5 & 1.0 & 30.12 & 0.15 & 0.98 & 6.11 & 0.03 \\
 3.5 & 3.0 & 136.55 & 0.36 & 9.19 & 8.89 & 0.07\\
\enddata
\tablecomments{See Table \ref{EnstatBASE} for definitions of quantities.}
\end{deluxetable*}

\begin{deluxetable*}{crrrrrr}
\tablecaption{
\label{EnstatSUBSTRU}
Encounter Statistics for model Q0.75D1.6}
\tablehead{
\colhead{$\log(M_c/M_{\odot})$} &
\colhead{$\Sigma_c$} &
\colhead{$N_{\rm enc}$} &
\colhead{$\sigma_{\rm Poisson}$} &
\colhead{$\sigma_{\rm sample}$} &
\colhead{$v_{\infty}^{\rm median}$} &
\colhead{$\sigma_r$} \\
\colhead{} &
\colhead{[g cm$^{-2}$]} &
\colhead{} &
\colhead{} &
\colhead{} &
\colhead{ [km s$^{-1}$]} &
\colhead{}
}
\startdata
 1.5 & 0.1 & 2.69 & 0.04 & 0.00 & 2.13 & 0.01 \\
 1.5 & 0.5 & 8.01 & 0.06 & 0.01 & 3.23 & 0.01\\
 1.5 & 1.0 & 12.86 & 0.08 & 0.02 & 3.66 & 0.01\\
 2.0 & 0.1 & 7.56 & 0.07 & 0.03 & 3.02 & 0.01\\
 2.0 & 0.5 & 22.05 & 0.14 & 0.12 & 4.46 & 0.01 \\
 2.0 & 1.0 & 34.47 & 0.17 & 0.18 & 4.99 & 0.01 \\
 2.5 & 0.1 & 14.03 & 0.10 & 0.12 & 3.57 & 0.01\\
 2.5 & 0.5 & 60.88 & 0.23 & 0.66 & 5.25 & 0.01\\
 2.5 & 1.0 & 132.85 & 0.50 & 5.88 & 6.76 & 0.04\\
 3.0 & 0.1 & 34.22 & 0.19 & 1.45 & 4.58 & 0.04 \\
 3.0 & 0.5 & 136.45 & 0.43 & 4.86 & 6.53 & 0.04\\
 3.0 & 1.0 & 275.40 & 0.71 & 16.94 & 8.52 & 0.06\\
 3.5 & 0.1 & 50.70 & 0.16 & 2.02 & 5.66 & 0.04\\
\enddata
\tablecomments{See Table \ref{EnstatBASE} for definitions of quantities.}
\end{deluxetable*}

\begin{deluxetable*}{crrrrrr}
\tablecaption{
\label{EnstatGAS}
Encounter Statistics for model GAS}
\tablehead{
\colhead{$\log(M_c/M_{\odot})$} &
\colhead{$\Sigma_c$} &
\colhead{$N_{\rm enc}$} &
\colhead{$\sigma_{\rm Poisson}$} &
\colhead{$\sigma_{\rm sample}$} &
\colhead{$v_{\infty}^{\rm median}$}&
\colhead{$\sigma_r$} \\
\colhead{} &
\colhead{[g cm$^{-2}$]} &
\colhead{} &
\colhead{} &
\colhead{} &
\colhead{ [km s$^{-1}$]} &
\colhead{}
}
\startdata
 1.5 & 0.1 & 2.79 & 0.05 & 0.01 & 2.24 & 0.02 \\
 1.5 & 0.5 & 13.08 & 0.12 & 0.04 & 3.12 & 0.01   \\
 1.5 & 1.0 & 23.42 & 0.17 & 0.07 & 3.78 & 0.01 \\
 1.5 & 3.0 & 53.34 & 0.30 & 0.20 & 4.92 & 0.01 \\
 2.0 & 0.1 & 3.70 & 0.05 & 0.02 & 2.91 & 0.01 \\
 2.0 & 0.5 & 19.62 & 0.14 & 0.11 & 4.29 & 0.01 \\
 2.0 & 1.0 & 35.84 & 0.21 & 0.23 & 5.20 & 0.01 \\
 2.0 & 3.0 & 134.21 & 0.51 & 1.61 & 6.55 & 0.01 \\
 2.5 & 0.1 & 4.97 & 0.06 & 0.06 & 3.59 & 0.02 \\
 2.5 & 0.5 & 25.96 & 0.17 & 0.46 & 5.48 & 0.02 \\
 2.5 & 1.0 & 57.29 & 0.27 & 0.66 & 6.74 & 0.01 \\
 2.5 & 3.0 & 250.26 & 0.70 & 4.70 & 9.27 & 0.02 \\
 3.0 & 0.1 & 5.47 & 0.07 & 0.13 & 4.26 & 0.03 \\
 3.0 & 0.5 & 29.04 & 0.18 & 0.74 & 6.98 & 0.03 \\
 3.0 & 1.0 & 88.33 & 0.32 & 2.91 & 8.53 & 0.03 \\
 3.0 & 3.0 & 318.47 & 0.77 & 27.18 & 10.79 & 0.09 \\
 3.5 & 0.1 & 6.40 & 0.06 & 0.25 & 5.21 & 0.04 \\
 3.5 & 0.5 & 46.90 & 0.17 & 2.96 & 8.54 & 0.06  \\
 3.5 & 1.0 & 99.66 & 0.28 & 9.34 & 9.67 & 0.09\\
 3.5 & 3.0 & 559.00 & 0.79 & 16.66 & 14.13 & 0.03 \\
 4.0 & 0.1 & 9.62 & 0.05 & 0.44 & 6.25 & 0.05 \\
 4.0 & 0.5 & 63.91 & 0.14 & 4.20 & 9.30 & 0.07 \\
 4.0 & 1.0 & 95.96 & 0.22 & 3.08 & 11.94 & 0.03\\

\enddata
\tablecomments{See Table \ref{EnstatBASE} for definitions of quantities.}
\end{deluxetable*}

\subsection{Base Case (Q0.75D2.2)}

We first describe the results of our base case, model Q0.75D2.2, and then in subsequent sections describe how the results change for the other models. We report the quantitative results for all models in Tables \ref{EnstatBASE} -- \ref{EnstatGAS}.

\subsubsection{Distributions of Encounter Velocities and Impact Parameters}

\begin{figure}
\plotone{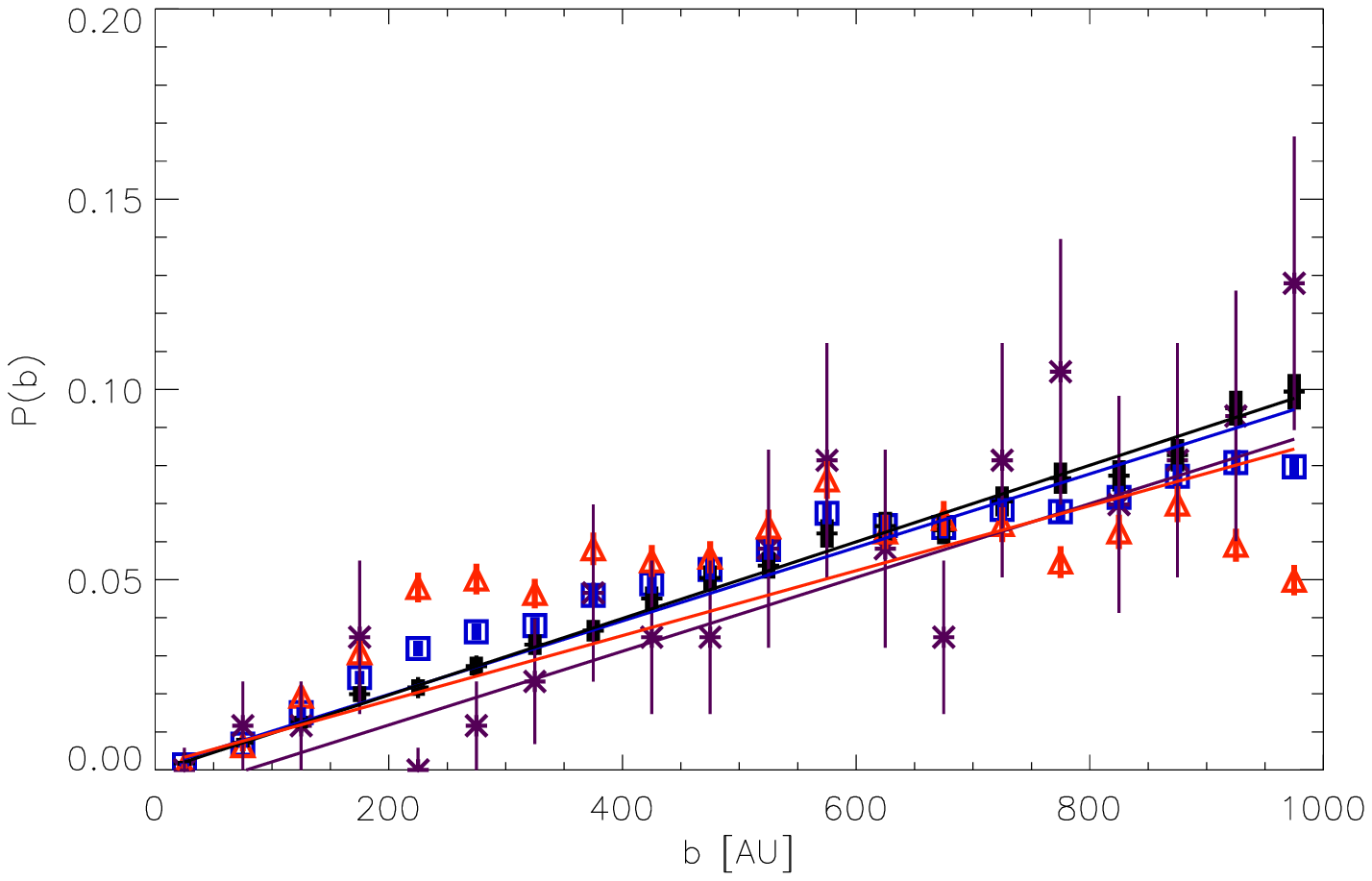}
\plotone{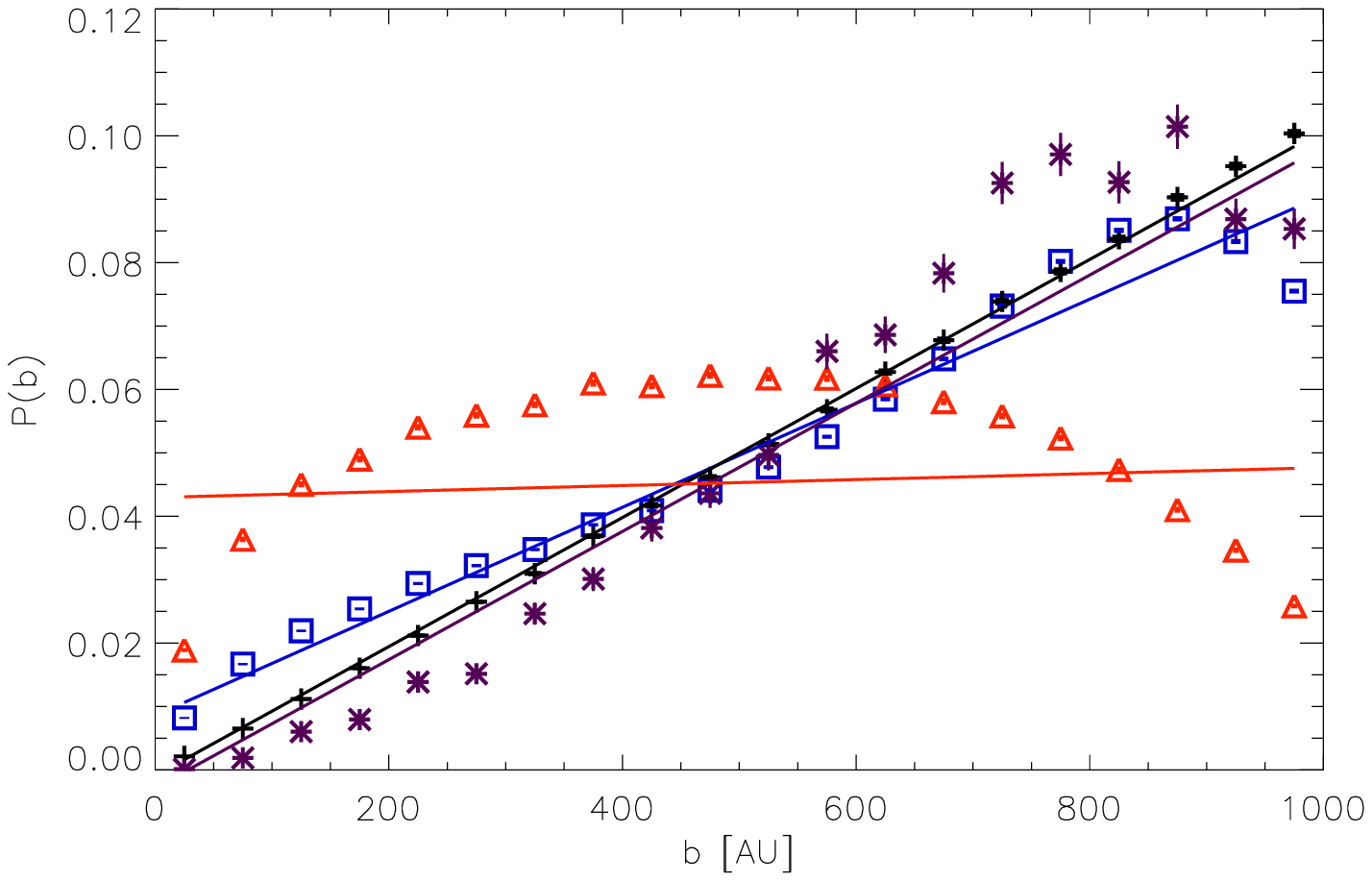}
\caption{Distribution of impact parameters for run Q0.75D2.2, with $M_c = 10^{1.5}$ $M_{\odot}$, $\Sigma_c = 0.1$ g cm$^{-2}$ (top), and for $M_c = 10^{4.5}$  $M_{\odot}$, $\Sigma_c = 1.0$ g cm$^{-2}$ (bottom). Black plus signs show the $1+1$ encounters, orange triangles the $1+1+1$ encounters, purple stars $1+2$ encounters, and blue squares are the sum of all interactions with a well-defined impact parameter. In all cases data points show the results of the simulation, with error bars indicating the $1\sigma$ Poisson error, and lines show linear best fits to the data.}
\label{fig:btest}
\end{figure}

Figure \ref{fig:btest} shows the distributions of impact parameter for two example runs, one at low and one at high surface density. We find that the distribution of impact parameters follows equation \eqref{bdistri}, $p(b) \propto b$, very closely for many of our cases, and in all cases the distribution of $1+1$ events follows a linear trend. We find a deviation from linearity with the $1+1+1$ events. This is not terribly surprising since we have made a rather large assumption that we can reliably describe a 3-body event as three 2-body events. As 3-body interactions become more prevalent with higher mass (for reasons to be discussed later) the deviation of the overall distribution from linearity tends to increase (blue in the figures). However, even for the highest surface density cases we consider, the overall distribution of impact parameters summed over all $1+1$, $1+2$, and $1+1+1$ events remains reasonably linear.

\begin{figure}
\plotone{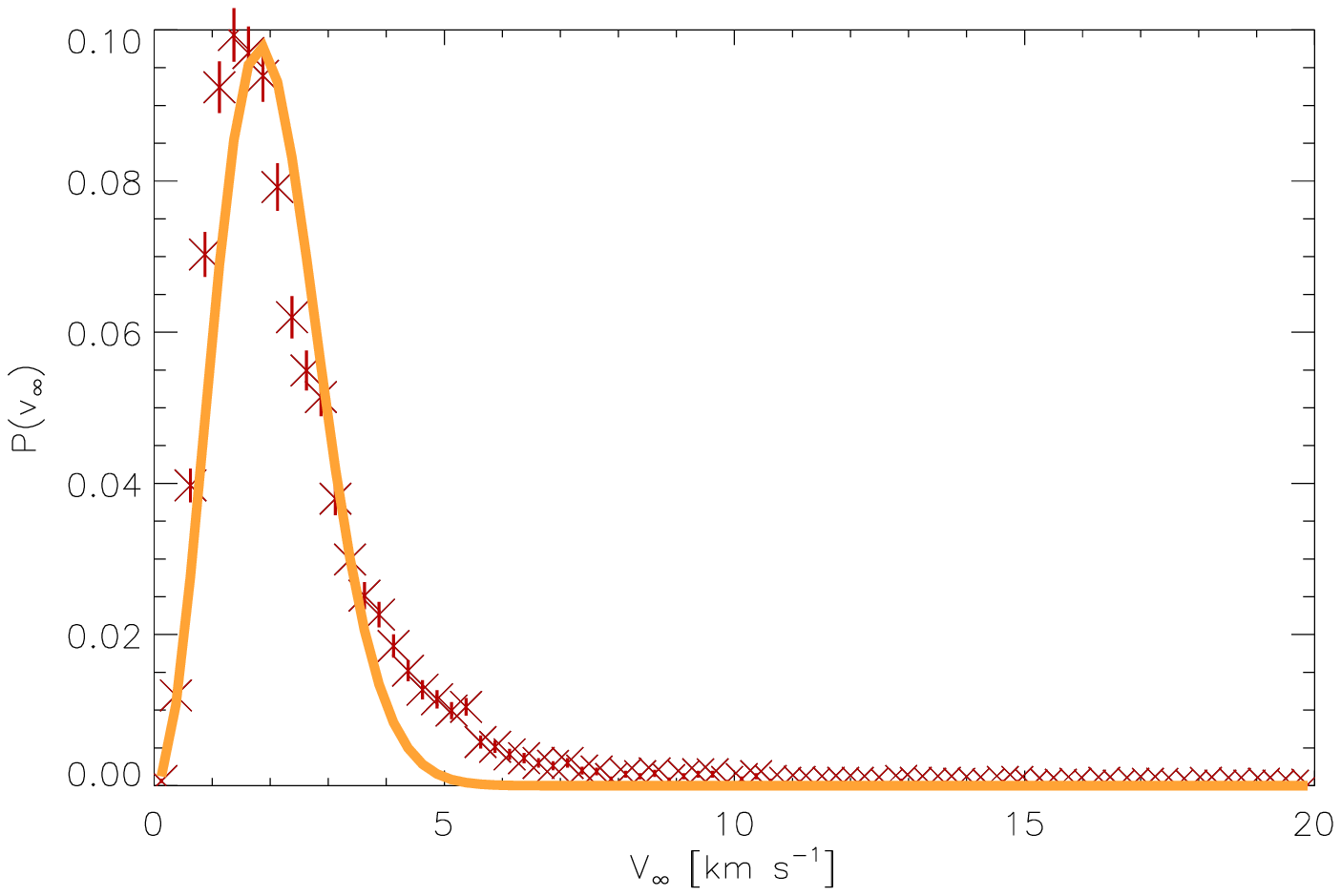}
\plotone{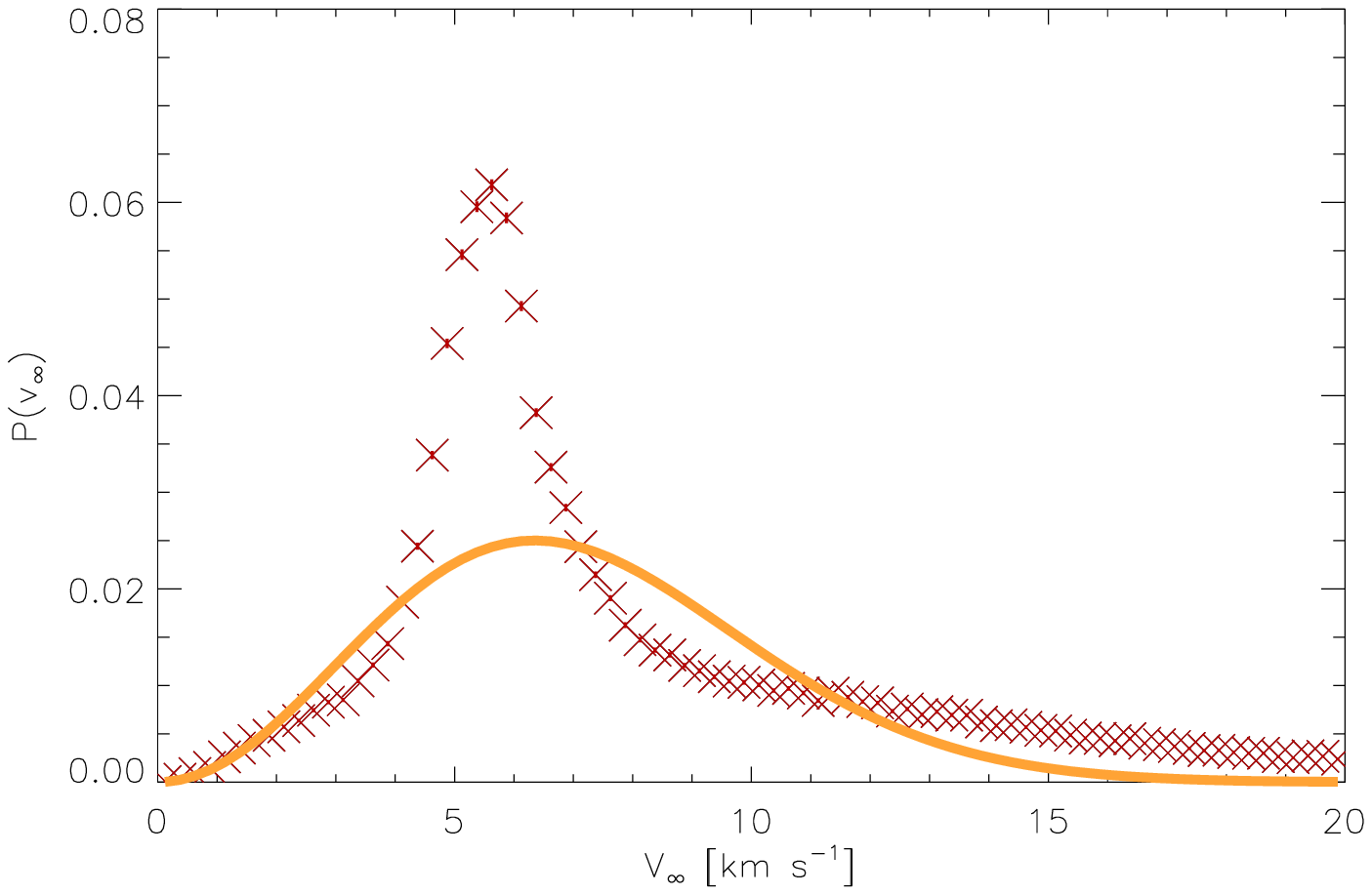}
\plotone{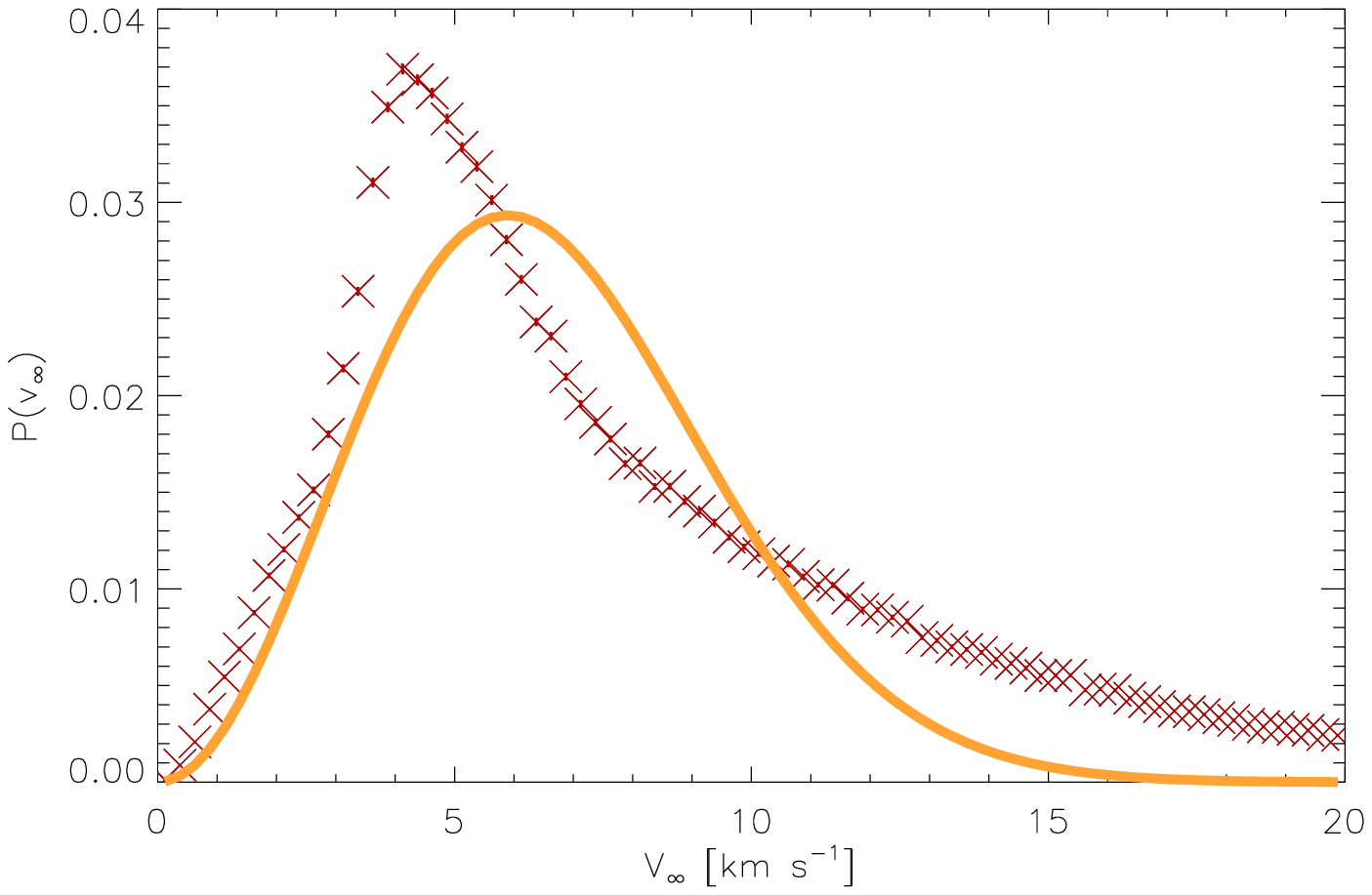}
\caption{Distributions of encounter relative velocities in model Q0.75D2.2, for the cases $(M_c,\Sigma_c)=(10^{1.5}, 0.1)$ (top), $(10^{3.5},3.0)$ (middle), and $(4.5,1.0)$ (bottom), where masses are in $M_\odot$ and $\Sigma_c$ in g cm$^{-2}$. Plus signs show data measured from the simulations, and lines show the best-fitting Maxwellian distribution. Typical reduced $\chi^2$ values are of order $100$, reflecting the poor fits. Notice the deviation is largest where the relative frequency of complex interactions is higher. 
\label{fig:vtest}} 
\end{figure}

While the distribution of impact parameters is close to what one would expect for a relaxed cluster, we find that the distribution of relative velocities is strongly non-Maxwellian in all our simulations. We show some examples of the distributions $p(v_\infty)$ from our simulations in Figure \ref{fig:vtest}. The deviation from Maxwellian is not surprising, given the correlated position-velocity distribution with which we begin, and that is observed in young clusters.

\subsubsection{Number of Encounters}

\begin{figure}
\plotone{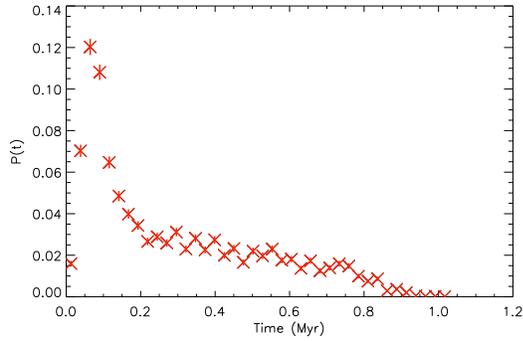}
\caption{Probability distribution of encounters in time in model Q0.75D2.2 for the case $M_c = 10^{1.5}$ $M_\odot$, $\Sigma_c=0.1$ g cm$^{-2}$. The crossing time for this model is $t_{\rm cr} = 0.15$ Myr, so the initial peak is less than one crossing time in duration.
\label{fig:tdistri}
}
\end{figure}

We now turn to the question of the number of encounters and their typical velocities as a function of $M_c$ and $\Sigma_c$. First note that nearly all of the events for these clusters occur in the first crossing time. Figure \ref{fig:tdistri} shows an example of the temporal distribution of encounters in one of our cases; other combinations of mass and surface density are similar or even more heavily weighted toward encounters occurring during the first crossing time. These results are similar to those of \citet{2012MNRAS.419.2448P}, who note that the stripping rate of planets from parent stars decreases with time, and \citet{2012MNRAS.427..637P}, who find that the surface density decreases rapidly after one crossing time.

\begin{figure}
\plotone{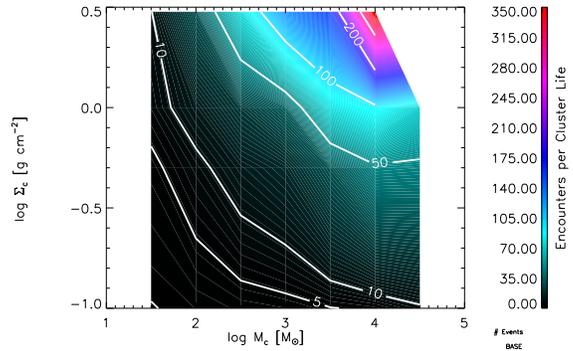}
\caption{The number of encounters a typical solar mass star can expect to experience after 5 crossing times in a fractal cluster, for the model $Q=0.75$ and $D=2.2$ as a function of the cluster mass and surface density.}
\label{fig:EcontQ075D22}
\end{figure}

For a relaxed, bound cluster, the mean number of encounters per star per crossing time (and thus over the cluster's entire life for a dispersing cluster) is a function of the cluster surface density alone, and does not depend on the mass \citep{Dukes}. We find that this is not the case for unrelaxed fractal clusters. Figure \ref{fig:EcontQ075D22} shows the number of encounters per solar mass star as a function of the cluster mass and surface density; clearly the number increases with both mass and surface density.

\begin{figure}
\plotone{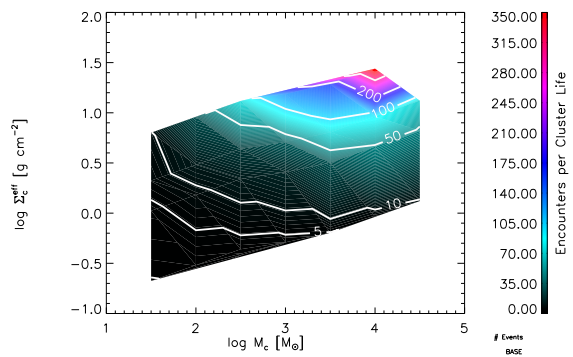}
\caption{Same as figure \ref{fig:EcontQ075D22} but with the effective surface density $\Sigma_{c,\rm eff}$.}
\label{fig:EconteffQ075D22}
\end{figure}

We can understand this result by realizing that the manner in which our fractal clusters are generated leads to an implicit dependence of the surface density on the mass of the cluster. This dependence arises because, although the mean surface density of the cluster averaged over its entire face is mass-independent, as the cluster mass increases at constant $\Sigma_c$ and $D$ the stars become packed into smaller and smaller substructures. In the Appendix we derive an expression for the effective surface density $\Sigma_{c,\rm eff}$ as a function of $M_c$ and $D$. Figure \ref{fig:EconteffQ075D22} shows the same data as Figure \ref{fig:EcontQ075D22}, but plotted using this effective surface density rather than the nominal one. As shown in the Figure, the number of encounters is in fact nearly independent of $M_c$ and fixed $\Sigma_{c,\rm eff}$.

One should regard this result with caution, since it is not clear that it remains valid for real clusters, which may or may not be truly fractal in their stellar distributions. Any attempt to define either $\Sigma_c$ or the volume density $\rho_c$ for a fractal cluster necessarily requires specifying an averaging scale over which the quantity is to be measured. $\Sigma_{c,\rm eff}$ is best considered as the surface density obtained by a process which averages over the initial clumps of the substructure, as opposed to the entire cluster. 
Alternately, one could envision $\Sigma_{c,\rm eff}$ as being closer to a mass-weighted surface density, as opposed to $\Sigma_c$, which is an area-weighted surface density.
 Our result simply shows that, if clusters are fractal, then more massive ones will produce more encounters than one might guess from their surface densities averaged over large scales. This is because in a fractal cluster the surface density increases as one averages over smaller and smaller scales in the vicinity of individual stars.

It is also interesting to compare our measurements to the results of a naive analytic estimate. For a uniform, spherical, virialized cluster of mass $M_c$ and surface density $\Sigma_c$, the expected number of encounters with impact parameter $b$ or less in a single crossing time is
\begin{equation}
N_{\rm enc,exp} \approx 2 \pi b^2 \frac{\Sigma_c}{\bar{m}} = 1.2 b_3^2 \Sigma_0,
\end{equation}
where $b_3=b/10^3$ AU, $\Sigma_0 = \Sigma_c/1$ g cm$^{-2}$, and we have used the mean stellar mass from our chosen IMF. Clearly the actual number of encounters we measure exceeds this value by a large margin.

\subsubsection{Encounter Velocities}

\begin{figure}
\plotone{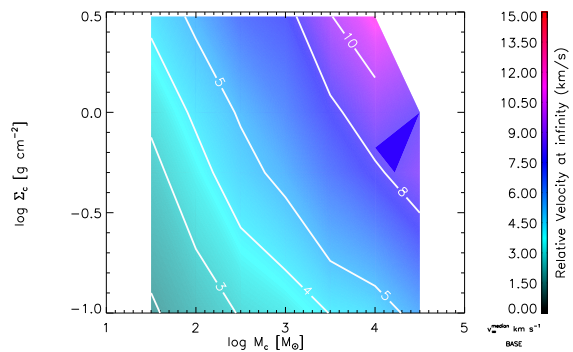}
\caption{Median $v_{\infty}$ as a function of $M_c$ and $\Sigma_c$ for run Q0.75D2.2.}
\label{fig:Q075D22vcont}
\end{figure}

\begin{figure}
\plotone{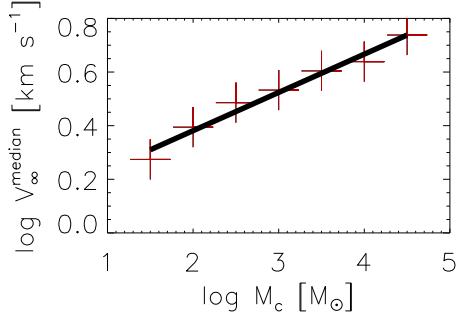}
\caption{Median $v_{\infty}$ as a function of the cluster mass at a fixed surface density $\Sigma_c = 0.1$ g cm$^{-2}$ in run Q0.75D2.2. The crosses are the simulations results and the line is the best linear fit, which has slope $0.15$.
\label{fig:Q075D22S01vtest}
}
\end{figure}

The typical encounter velocity also depends on $M_c$ and $\Sigma_c$. Since we found that the distribution of relative velocities was non-Maxwellian, rather than reporting a velocity dispersion we instead compute the median $v_{\infty}$ of the encounters as a function of $M_c$ and $\Sigma_c$. We plot the result in Figure \ref{fig:Q075D22vcont}. The dependence on $M_c$ and $\Sigma_c$ is somewhat unexpected. For a relaxed cluster, the stellar velocity dispersion obeys
\begin{equation}
\label{sigv}
\sigma_v \propto (M_c \Sigma_c)^{0.25}.
\end{equation}
For our non-relaxed clusters, we do see a general increase in the relative velocity with mass and surface density, as predicted by equation \eqref{sigv}. The increase is not as fast as expected, however. Figure \ref{fig:Q075D22S01vtest} shows $v_{\infty}$ versus $M_c$ at $\Sigma_c = 0.1$ g cm$^{-2}$ -- effectively a horizontal cut through Figure \ref{fig:Q075D22vcont}. The slope of the best-fit line is approximately $0.15$ rather than $0.25$, and this is true for each of the other surface density bins as well. As with the overall non-Maxwellian distribution, this slower than expected increase is a result of the correlated positions and velocities.

\subsection{Unbound Case (Q1.25D2.2)}

\begin{figure}
\plotone{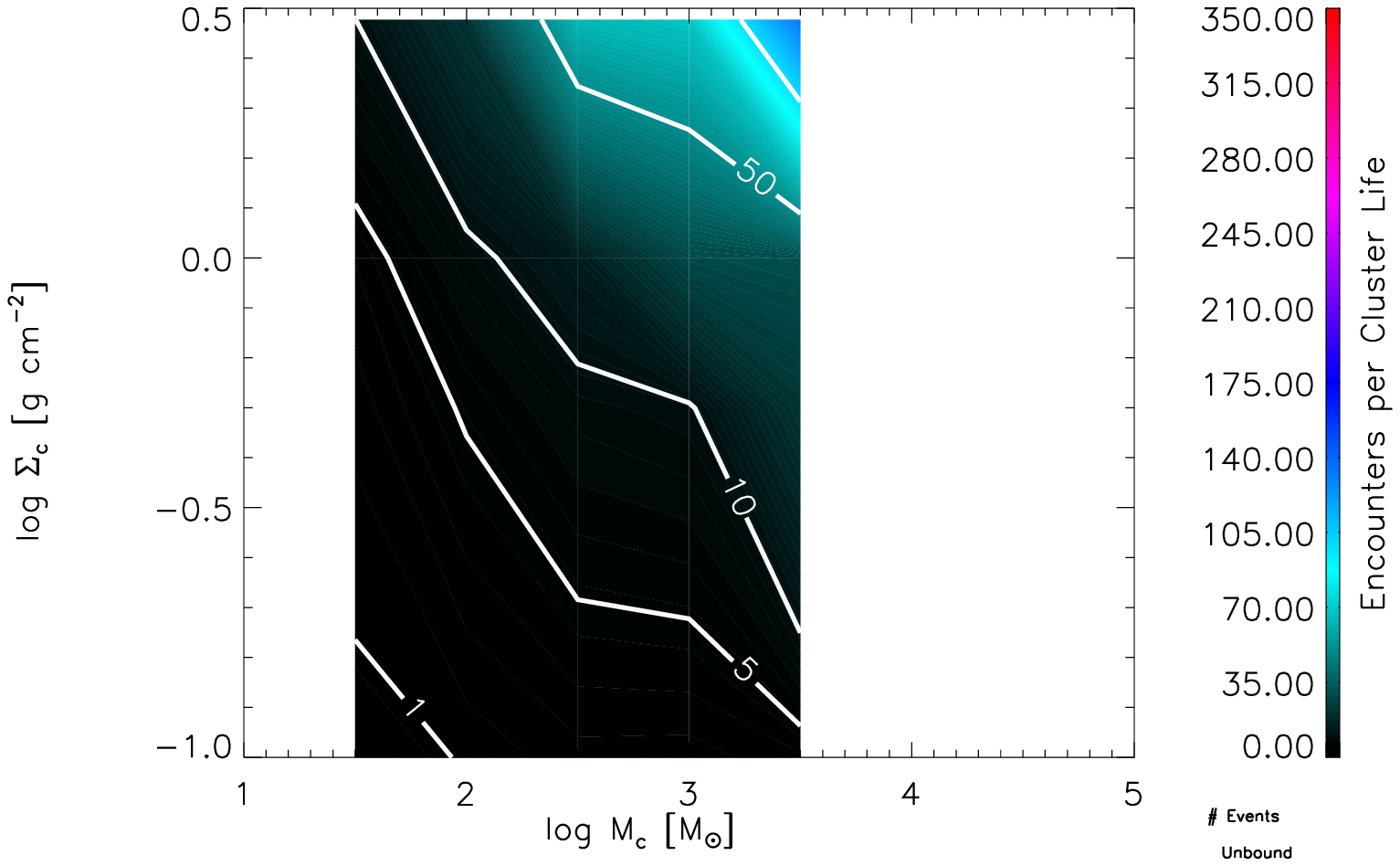}
\plotone{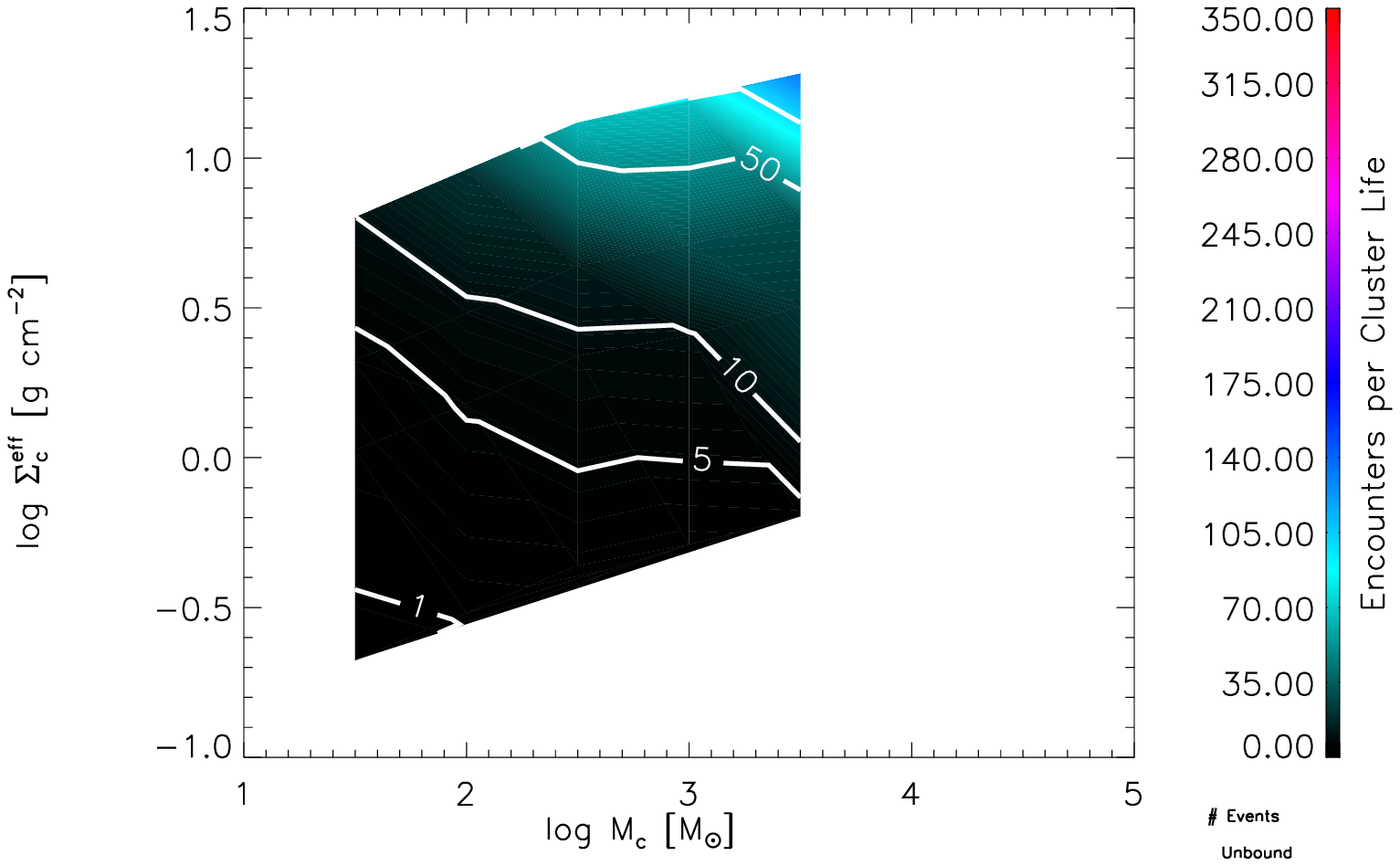}
\caption{Average number of encounters for a solar mass star in an unbound cluster (model Q1.25D2.2) as a function of $M_c$ and $\Sigma_c$ (top), $\Sigma_{c,\rm eff}$ (bottom).
\label{fig:Q125D22Encounter}
}
\end{figure}

\begin{figure}
\plotone{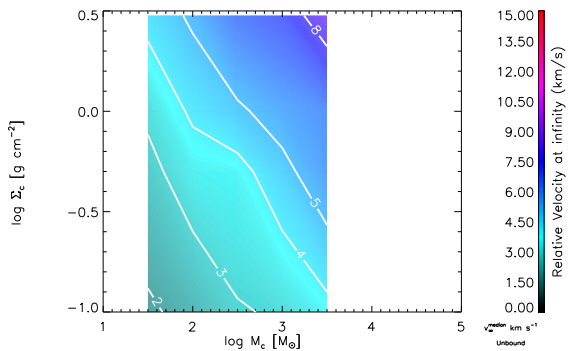}
\caption{Same as figure~\ref{fig:Q075D22vcont} but for $Q=1.25$ and $D=2.2$.
\label{fig:Q125D22Vcont}
}
\end{figure}

In the case of an unbound cluster, the shapes of the distributions of $b$ and $v_\infty$ are quite similar to those found in the base case Q0.75D2.2, so we do not discuss them further. The number and median velocity of encounters, however, differ from the base case in interesting ways. Figure \ref{fig:Q125D22Encounter} shows the mean number of encounters per Solar-type star for the case of an unbound cluster. As expected there are fewer encounters in this model than there are in the case where at least some of the cluster remains bound. Figure~\ref{fig:Q125D22Vcont} shows the the median $v_{\infty}$. Rather surprisingly the velocities tend to actually be slower in this model than in the case of $Q=0.75$, despite the fact that the initial velocities are larger at the same $M_c$ and $\Sigma_c$ than in the base case. This behavior is a result of the initially correlated velocity structure. In the base case the cluster tends to relax towards equilibrium, destroying the velocity substructure in the process. Once the substructure is destroyed the velocity vectors are randomized, producing larger $v_{\infty}$ values than during the period when the velocity structure is coherent. In contrast, an unbound cluster tends to blow apart before substructure can be erased, leading us to a lower median velocity for those encounters that do occur.

\subsection{High Substructure (Q0.75D1.6)}

\begin{figure}
\plotone{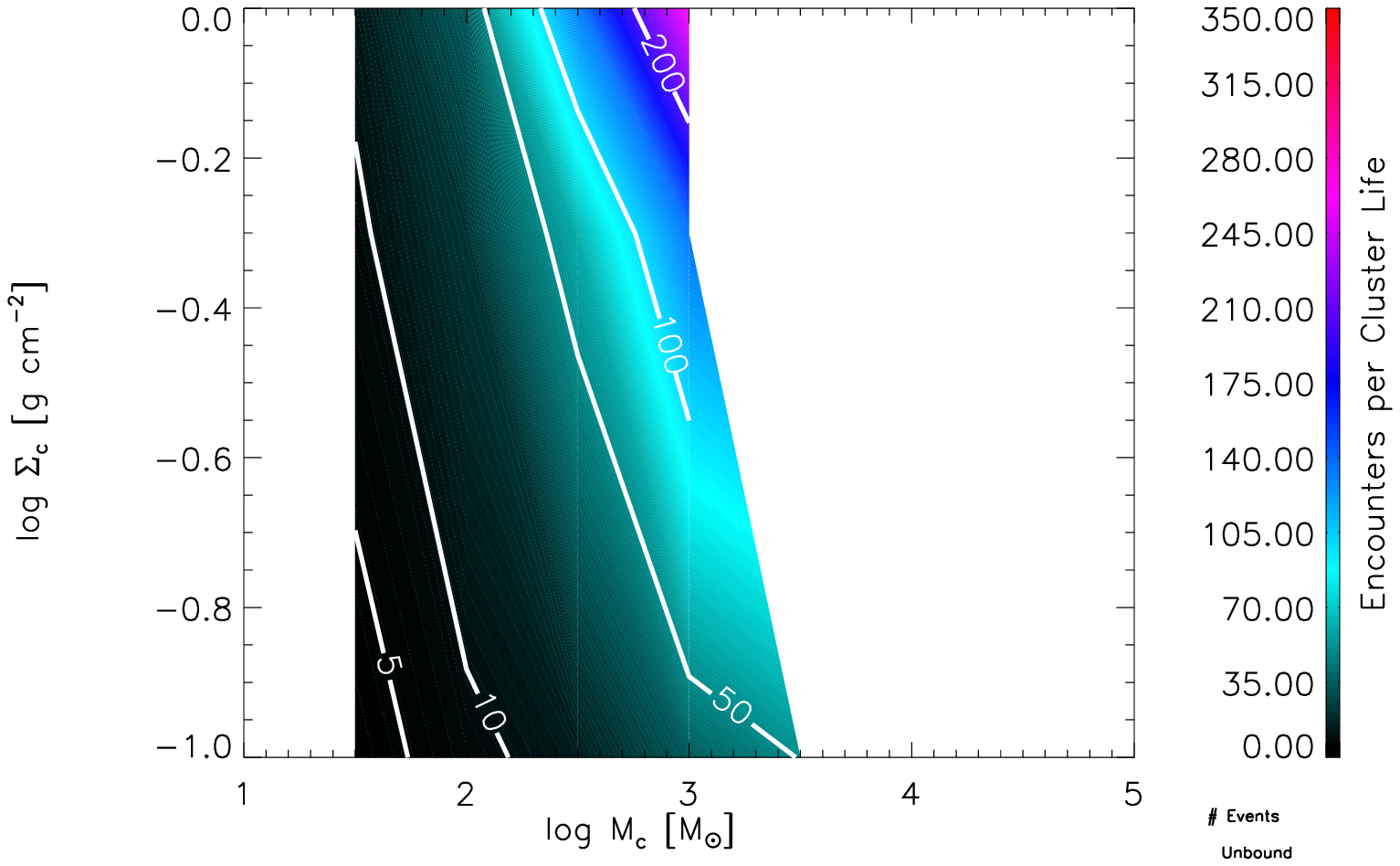}
\plotone{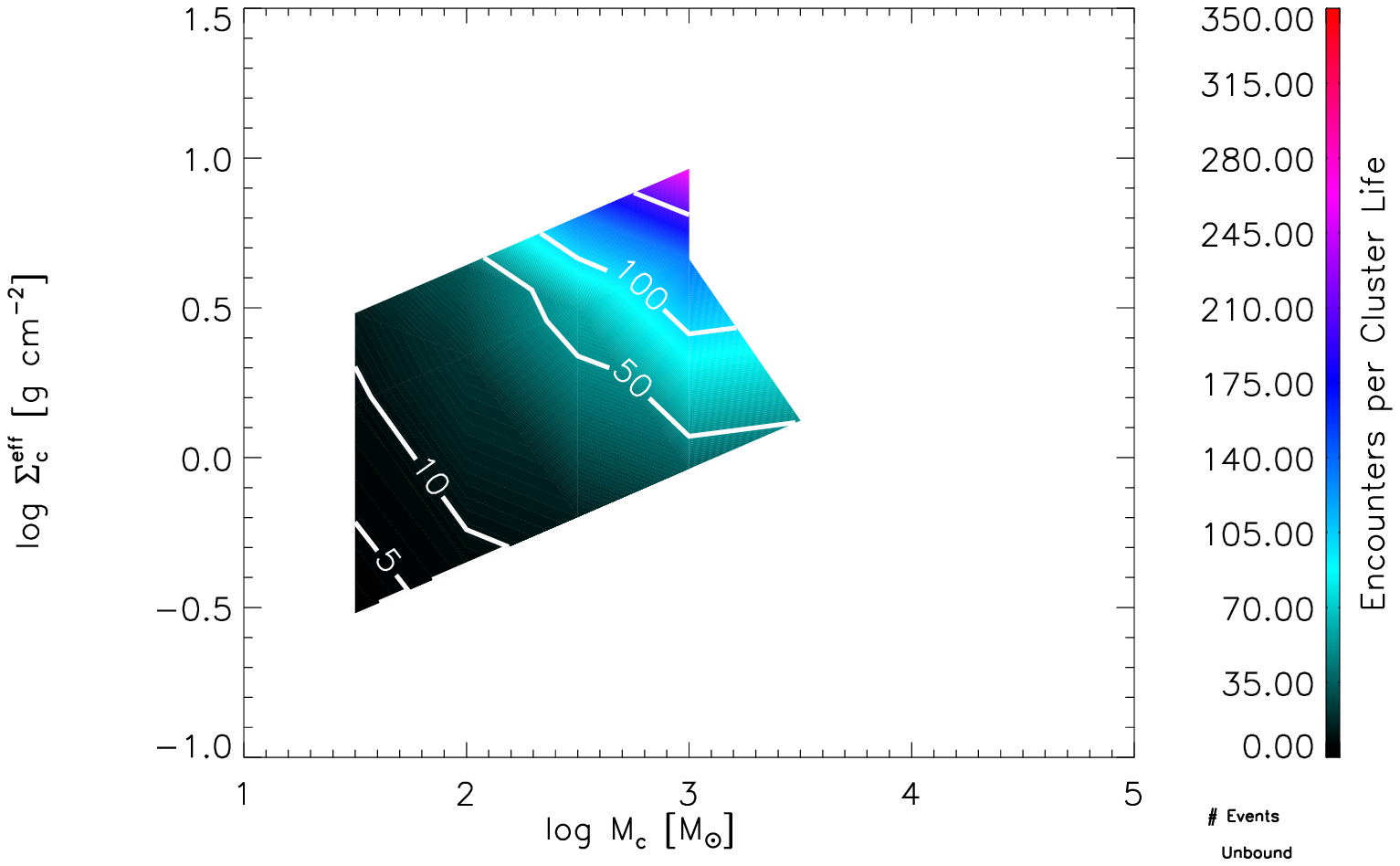}
\caption{Same as figure~\ref{fig:EcontQ075D22} but for $D=1.6$.
\label{fig:D16Econt}} 
\end{figure}

\begin{figure}
\plotone{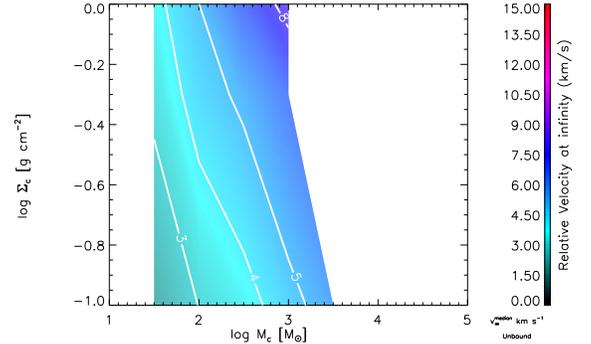}
\caption{Same as figure~\ref{fig:Q075D22vcont} but for $D=1.6$.
\label{fig:D16Vcont}
}
\end{figure}

The case of high substructure is the most extreme case we study in this paper, although it turns out qualitatively to be similar to the base case. Figures~\ref{fig:D16Econt} and~\ref{fig:D16Vcont} show the encounter statistics for this case. Both the number of encounters and the relative velocity are higher for this case than for the base case. This is consistent with the interpretation of $D$ as an increase in the effective surface density. However when we plot the number of encounters with $\Sigma_{c, \rm eff}$ the number of encounters in the $D=1.6$ cluster is typically higher than the corresponding point in the base case, and the contours are still largely vertical rather than horizontal. Thus our model for $\Sigma_{c, \rm eff}$ is not fully capturing the increase in encounters that occurs for very highly substructured clusters containing large numbers of stars.

\subsection{Gas Case (Q0.3D2.2)}

Since our Gas runs are extremely subvirial, we expect these clusters to relax and destroy substructure extremely quickly. Figure \ref{fig:GasPT} shows the temporal distribution of encounters for this model, which is consistent with this expectation. For the first crossing time, there is a highly elevated encounter rate as stars fall toward the center of the potential well and interact. After this they revirialize and the encounter rate drops and becomes roughly constant. Once gas is removed after four crossing times, the cluster disperses and the encounter rate drops to very small values.

\begin{figure}
\plotone{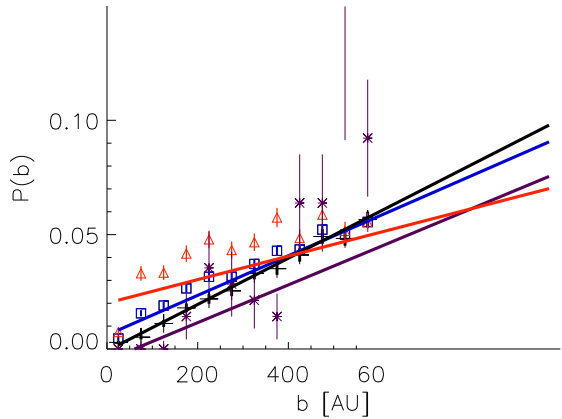}
\caption{The distribution of impact parameters for the GAS case with $M_c = 10^{3.0} M_{\odot}$ and $\Sigma_c = 0.5$ g cm$^{-2}$. Here we include only events which occur between one and four crossing times (the relaxed phase of our model). Black crosses are 1+1 events and purple stars are 2+1 events.
\label{fig:Relaxeddistribution}}
\end{figure}

\begin{figure}
\plotone{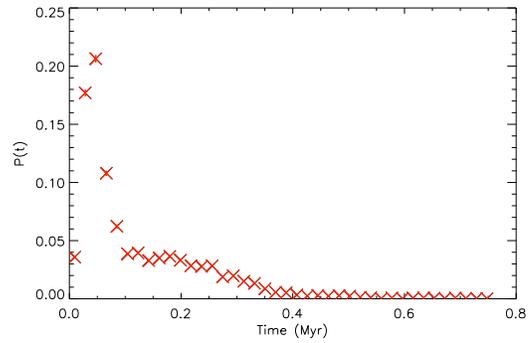}
\caption{The temporal distribution of encounters for a typical gas case. This particular model has $M_c = 10^{2.5} m_{\odot}$, $\Sigma_c = 0.5$ g cm$^{-2}$, so the crossing time is $t_{\rm cr} = 0.078$ Myr. Gas removal occurs at $4t_{\rm cr} = 0.31$ Myr.
\label{fig:GasPT}
}
\end{figure}

The results of \citet{2012MNRAS.424..272P} suggest that the distribution function of impact parameters (equation \eqref{bdistri}) could be significantly altered even in relaxed clusters due to the presence of intermediate separation binaries. To investigate this possibility, in Figure \ref{fig:Relaxeddistribution} we shows the ensemble distribution of impact parameters for our Gas model, considering only encounters that occur at times from $1-4t_c$. During this phase the cluster is well-relaxed, since it is old enough to have lost most of its initial substructure, but we have not yet removed the confining gas potential. As the Figure shows, the distribution function still follows $P(b) \propto b$ for 1+1 interactions, although some evidence of stochastic behavior is observed for the 2+1 encounters. We have typically seen such deviations for 2+1 encounters (such as figure \ref{fig:btest}), and this is probably more due to the difficulty of assigning orbital elements when a binary interacts with the single star, as opposed to the conditions within the cluster more broadly.

\begin{figure}
\plotone{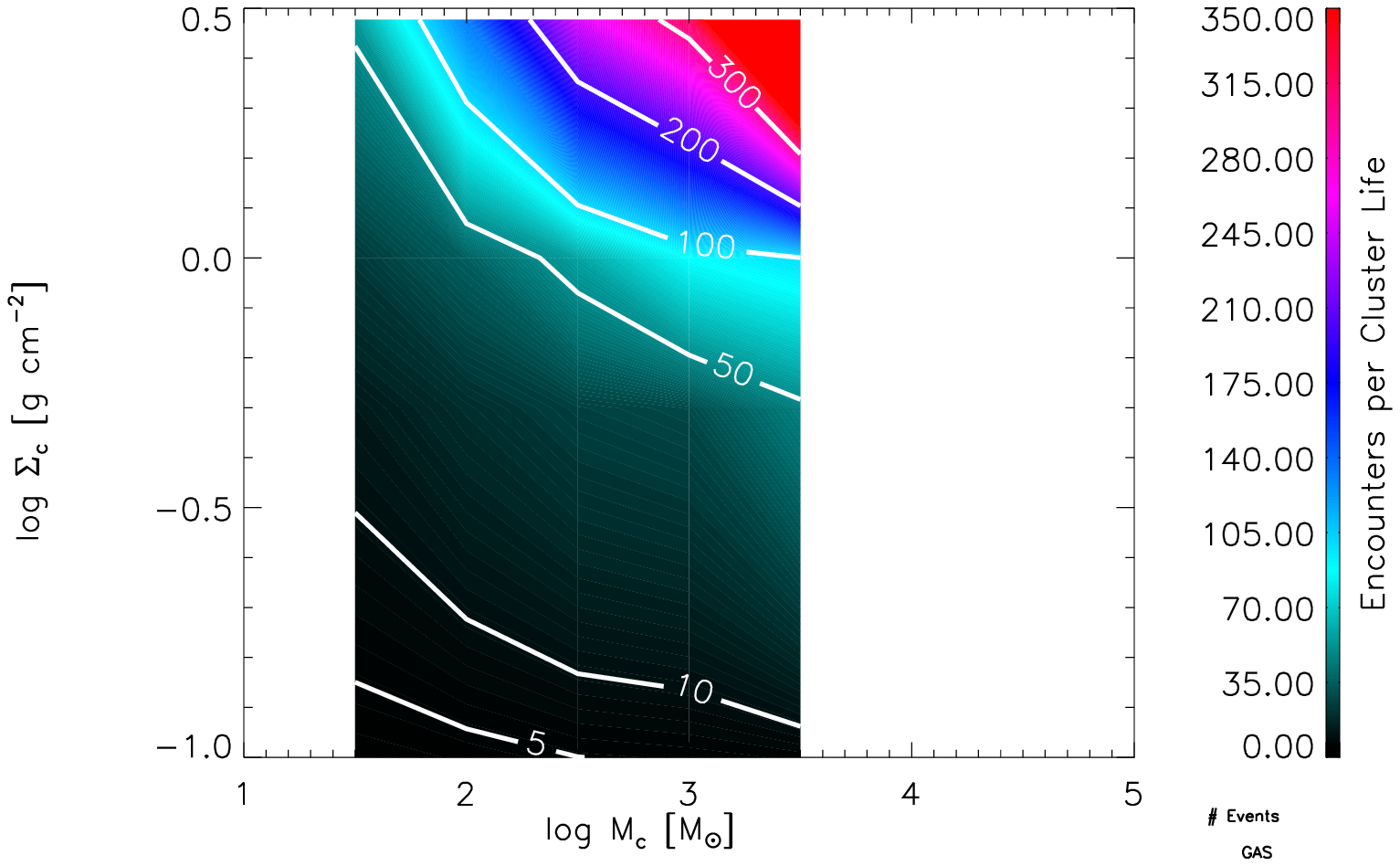}
\plotone{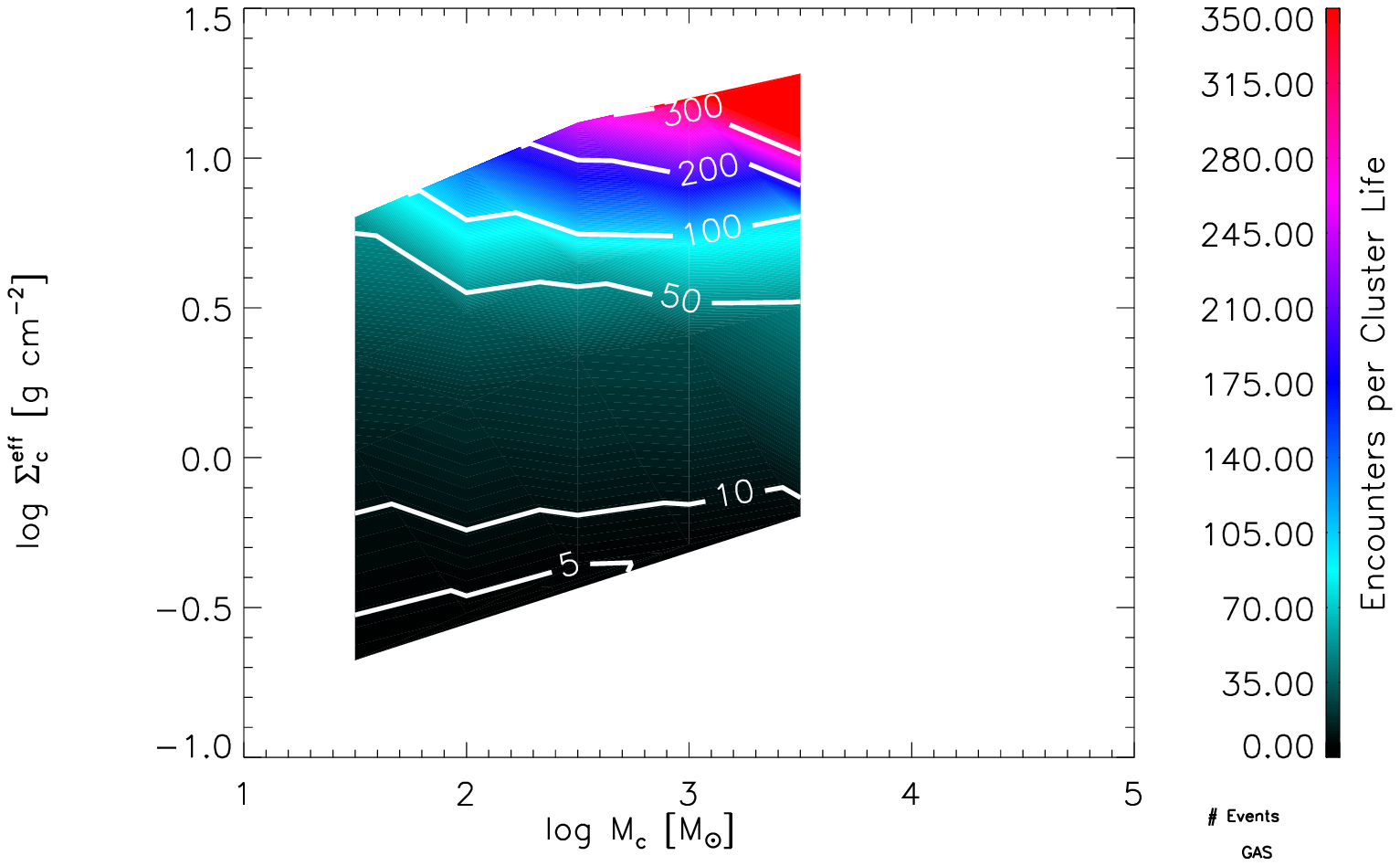}
\caption{Same as figure~\ref{fig:EcontQ075D22} but for the gas case.
\label{fig:GASEcont}}
\end{figure}

\begin{figure}
\plotone{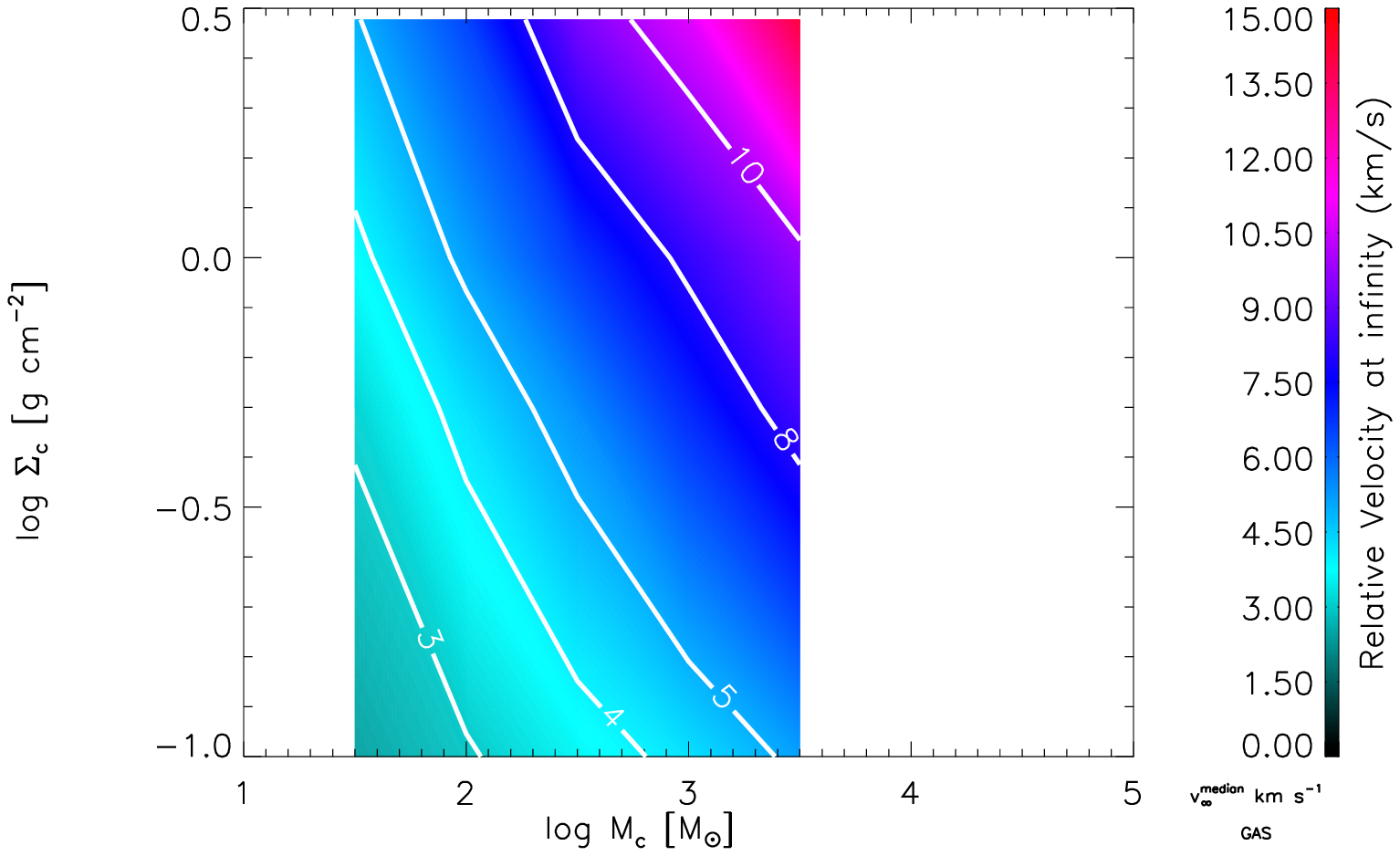}
\caption{Same as figure~\ref{fig:Q075D22vcont} but for the gas case.
\label{fig:GASVcont}
}
\end{figure}

\begin{figure}
\plotone{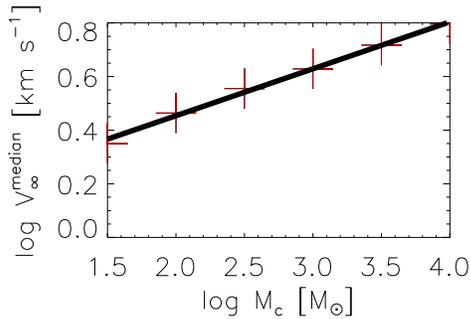}
\caption{The log of the median $v_{\infty}$ versus the log of $M_c$ at $\Sigma_c=0.1$ g cm$^{-2}$. The red crosses are our data points with the best-fit line drawn in black. The slope in this case is 0.17.}
\label{fig:GASvgrow}
\end{figure}

The number of encounters is shown in Figure \ref{fig:GASEcont}, and is larger than the number of encounters in the base case (shown in Figure \ref{fig:EcontQ075D22}), but only slightly -- certainly by less than the factor of 4 difference in times for which the cluster survives before dispersing. Contours of encounter number in the $(\Sigma_c, M_c)$-plane are somewhat flatter for the Gas case than the base case, and are much flatter in the plane of $(\Sigma_{c,\rm eff}, M_c)$. Comparing Figures \ref{fig:GASVcont} and \ref{fig:Q075D22vcont}, we see that median encounter velocities are larger in the Gas case than in the base case. Figure \ref{fig:GASvgrow} shows the results of taking a horizontal slice to see how the median encounter velocity behaves as a function of $M_c$ at fixed $\Sigma_c$. We find that the median encounter velocity is increasing more quickly with mass in the Gas case than the base case, but not quite as quickly as for a fully relaxed cluster.

We can summarize all of these results by saying that the Gas case represents a compromise between the case of a fully relaxed cluster and the substructured clusters we have considered thus far; the first crossing time, which produces a significant fraction of the encounters due to the elevated encounter rate during the relaxation phase, looks much like the substructured clusters. After a crossing time the stars revirialize, and the evolution from that point until gas expulsion looks like that in a dynamically relaxed cluster. Because a significant fraction of the encounters occur during the early, subvirial relaxation phase, the overall number of encounters and encounter velocity distribution ends up being intermediate between the substructured and fully-relaxed cases. A lifetime of four crossing times before gas expulsion is not long enough for the overall statistics to be dominated by the relaxed phase rather than the unrelaxed one.

\section{Discussion and Implications}
\label{sec:discussion}

\subsection{Error Analysis}
\label{sec:error}

How certain are the values of $N_{\rm enc}$ derived above, given the number of simulations we have run and the number of encounters they produce? This question requires some care. There are two distinct sources of error, and each dominates in a different regime of our simulations. One source of error is simply counting statistics on the total number of events at a given $(M_c,\Sigma_c)$. At low values of $M_c$ and $\Sigma_c$, even when we have a large number of runs, the total number of events recorded over all simulations may be small, producing a significant statistical error. The second source of error arises from our limited sampling of all possible realizations of fractal clusters at a given $(M_c,\Sigma_c)$. At large $(M_c,\Sigma_c)$ the number of events in a given run may be quite large, producing small Poisson errors, but the numbers of events may be quite different for different realizations at the same $(M_c,\Sigma_c)$. For example, we might have three realizations that produce 8,000, 10,000, and 12,000 events, respectively. In this case the Poisson error on each of these numbers is of order 100, much smaller than the difference between them, indicating that our error is dominated by our limited sampling of possible clusters at a given $(M_c, \Sigma_c)$. A reasonable estimate of the error in this regime is the standard deviations of the mean values for each run, neglecting the Poisson errors on each individual run. To interpolate between these two limits, we take the total error to be the quadrature sum of these two types of error. This is approximate, but should be roughly correct.


To make the above analysis precise,
let $N_{\rm run}$ be the number of simulations run for a given set of parameters $(M_c,\Sigma_c,Q,D)$. Let $S_i$ and $E_i$ be the number of Solar mass stars and encounters in the $i^{\rm th}$ realization, respectively. Then we define
\begin{eqnarray}
N_{\rm events} & =& \sum_{i=1}^{N_{\rm run}} E_i , \\
N_{*} & = & \sum_{i=1}^{N_{\rm run}} S_i ,
\end{eqnarray}
so that 
\begin{equation}
N_{\rm enc} = \frac{N_{\rm events}}{N_{*}}
\end{equation}
is our best estimate of the number of encounters per Sun-like star. From these definitions we can express the error on $N_{\rm enc}$ as 
\begin{equation}
\label{dNenc}
\delta N_{\rm enc} ^2 = \frac{\delta N_{\rm events}^2}{N_{*}^2} = \sigma_{\rm Poisson}^2 + \sigma_{\rm sample} ^2 ,
\end{equation}
where
\begin{equation}
\label{eq:sigmapoisson}
\sigma_{\rm Poisson} = \frac{\sqrt{N_{\rm events}}}{N_{*}},
\end{equation}
is the error due to counting statistics\footnote{Note that this expression assumes the Gaussian limit, and therefore becomes invalid when $N_{\rm events} \la 10$.} and
\begin{equation}
\label{eq:sigmasample}
\sigma_{\rm sample} = \frac{\left[\sum_{i=1}^{N_{\rm run}} \left(E_i-\bar{E} \right)^2/N_{\rm run}\right]^{1/2}}{N_{*}}
\end{equation}
is the error introduced by a lack of ability to fully sample the parameter space by having enough realizations of initial clusters. Here $\bar{E}$ is the mean number of events per run. We report $\sigma_{\rm Poisson}$ and $\sigma_{\rm sample}$ separately in Tables \ref{EnstatBASE} -- \ref{EnstatGAS}, along with the total relative error
\begin{equation}
\label{eq:sigmar}
\sigma_r = \frac{\delta N_{\rm enc}}{N_{\rm enc}}.
\end{equation}

\subsection{The Sun's Birth Environment}

One of the primary applications of the statistics we have measured is constraining the environment in which the Sun was born. To do so, we make use the the velocity-dependent cross sections for perturbing the orbits of the outer planets measured by \citet{Dukes} from their simulations. In deriving these values \citeauthor{Dukes} implicitly integrated over the distribution of impact parameters under the assumption that this distribution follows $p(b) \propto b$, and we have found that this is generally a good assumption. 

If $\sigma_{i}(v_{\infty})$ is the cross section for a particular event to occur for stars approaching with a particular relative velocity $v_\infty$, the expected number of occurrences of that event in a cluster within which there are expected to be $N_{\rm enc}$ encounters with impact parameter less than $b_{\rm max} = 1000$ AU is
\begin{equation}
\Lambda_i = \frac{N_{\rm enc}}{\pi b_{\rm max}^2} \frac{\int_0^{\infty}{\sigma_i (v_\infty)}v_\infty p(v_\infty) \, dv_\infty}{\int_0^{\infty} v_\infty p(v_\infty) \, dv_\infty}.
\label{pint}
\end{equation}
Assuming the events are Poisson in nature (i.e.~that they are independent), the probability of an occurrence is simply
\begin{equation}
P_i = 1 - \exp(-\Lambda_i).
\end{equation}
Following \citet{Dukes}, we examine the possibility that a close encounter with a passing star might excite one of the Jovian planets to a highly eccentric ($e > 0.1$) orbit\footnote{Alternately, since the Jovian planets are likely not fully formed during the early phases of evolution we are considering, we can regard these probabilities as the describing the chances that an encounter with another star might severely perturb the protoplanetary disk at the radii where the Jovian planets are located now.}. To evaluate $\Lambda_i$, we combine the measured values of $\sigma_i(v_\infty)$ from \citeauthor{Dukes} with velocity distributions $p(v_\infty)$ obtained in the previous section.

\begin{figure}
\plottwo{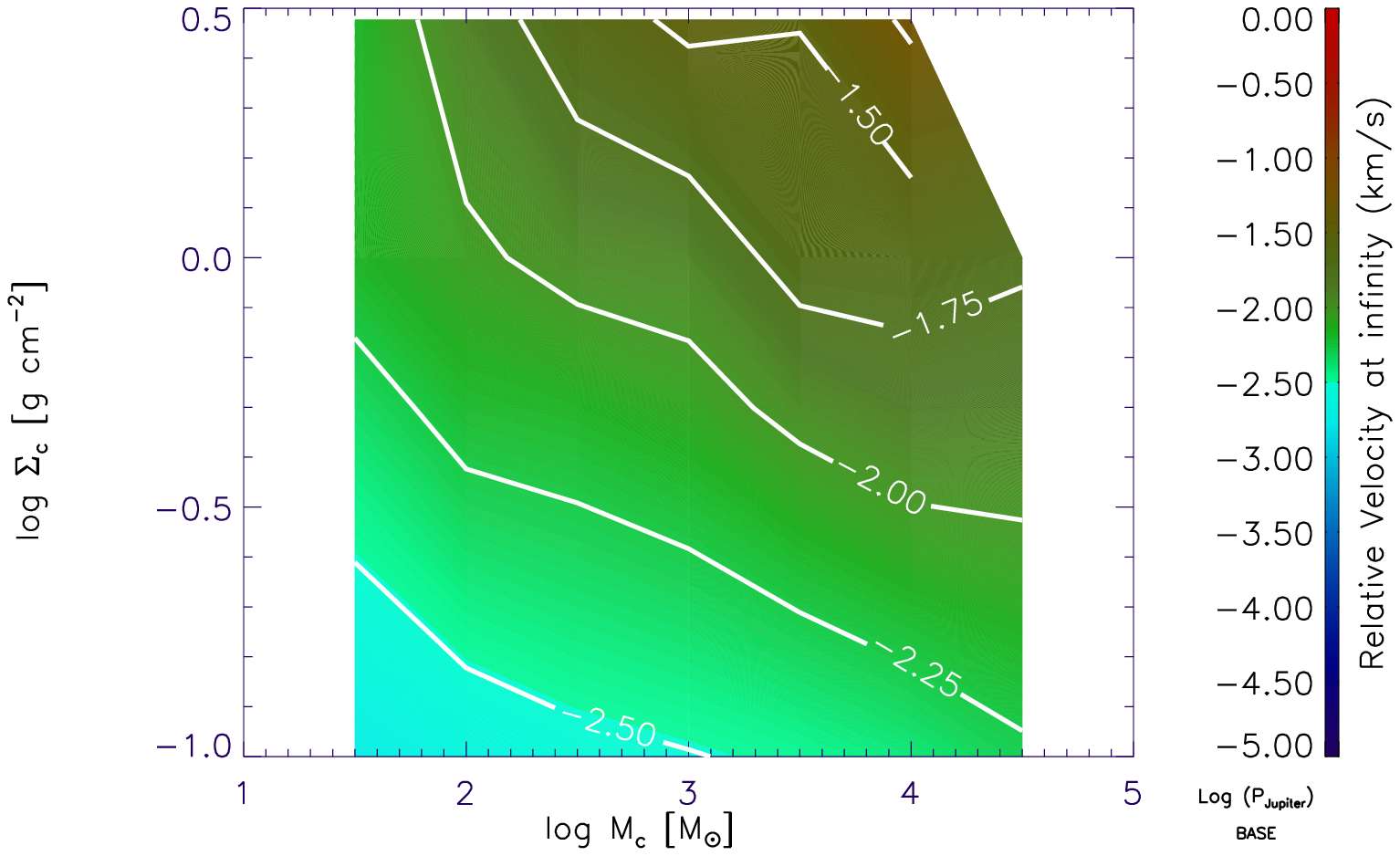}{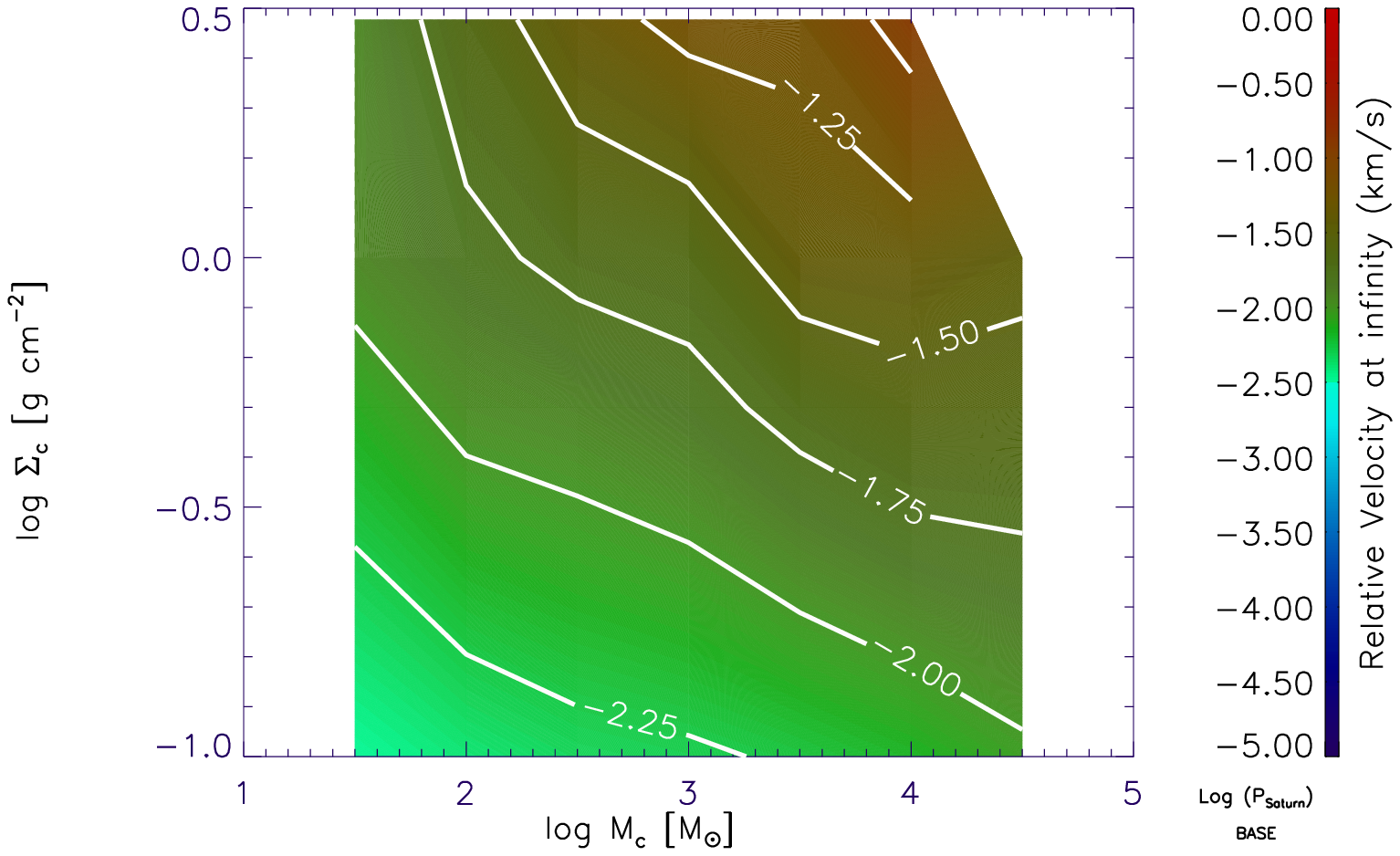}
\plottwo{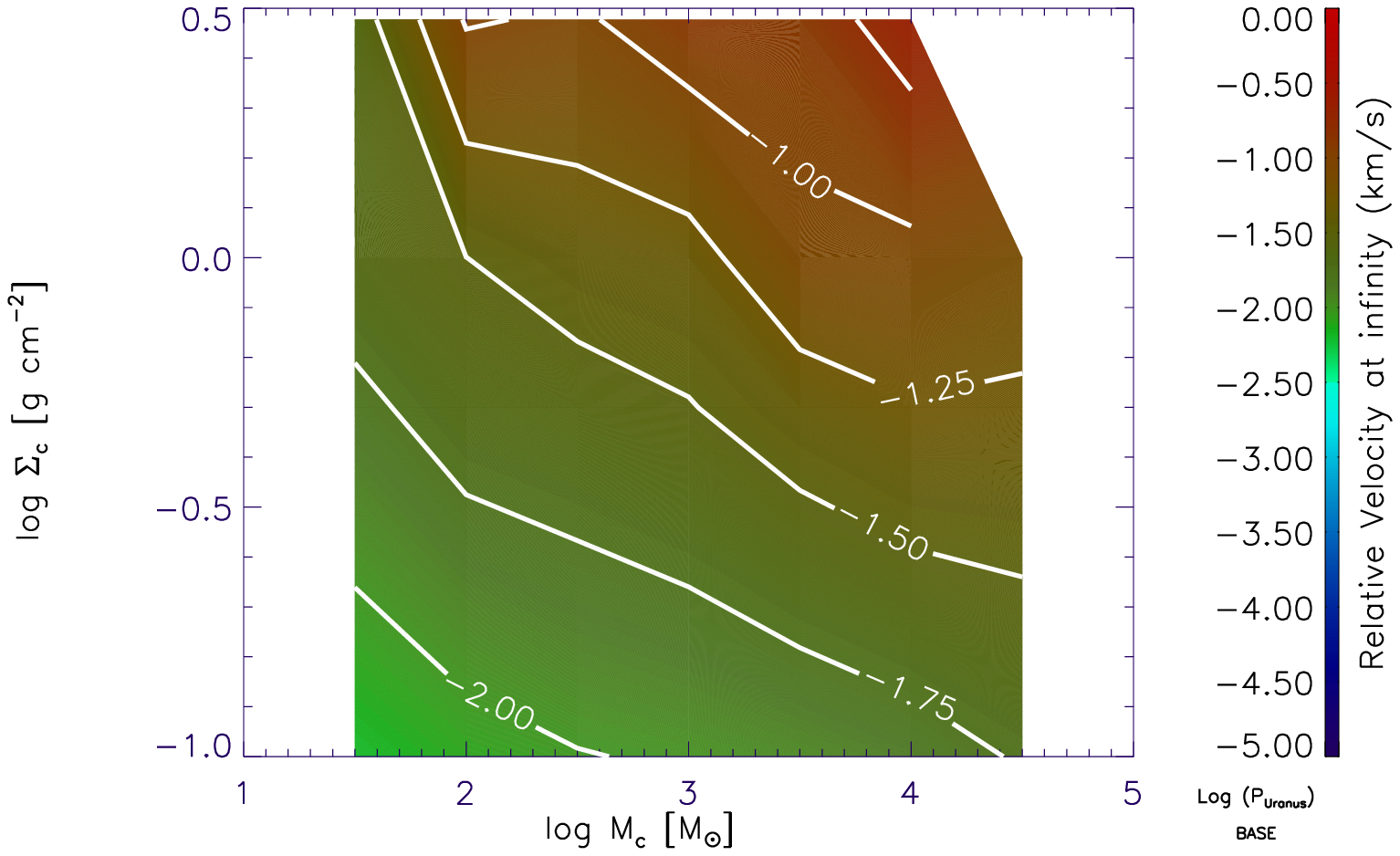}{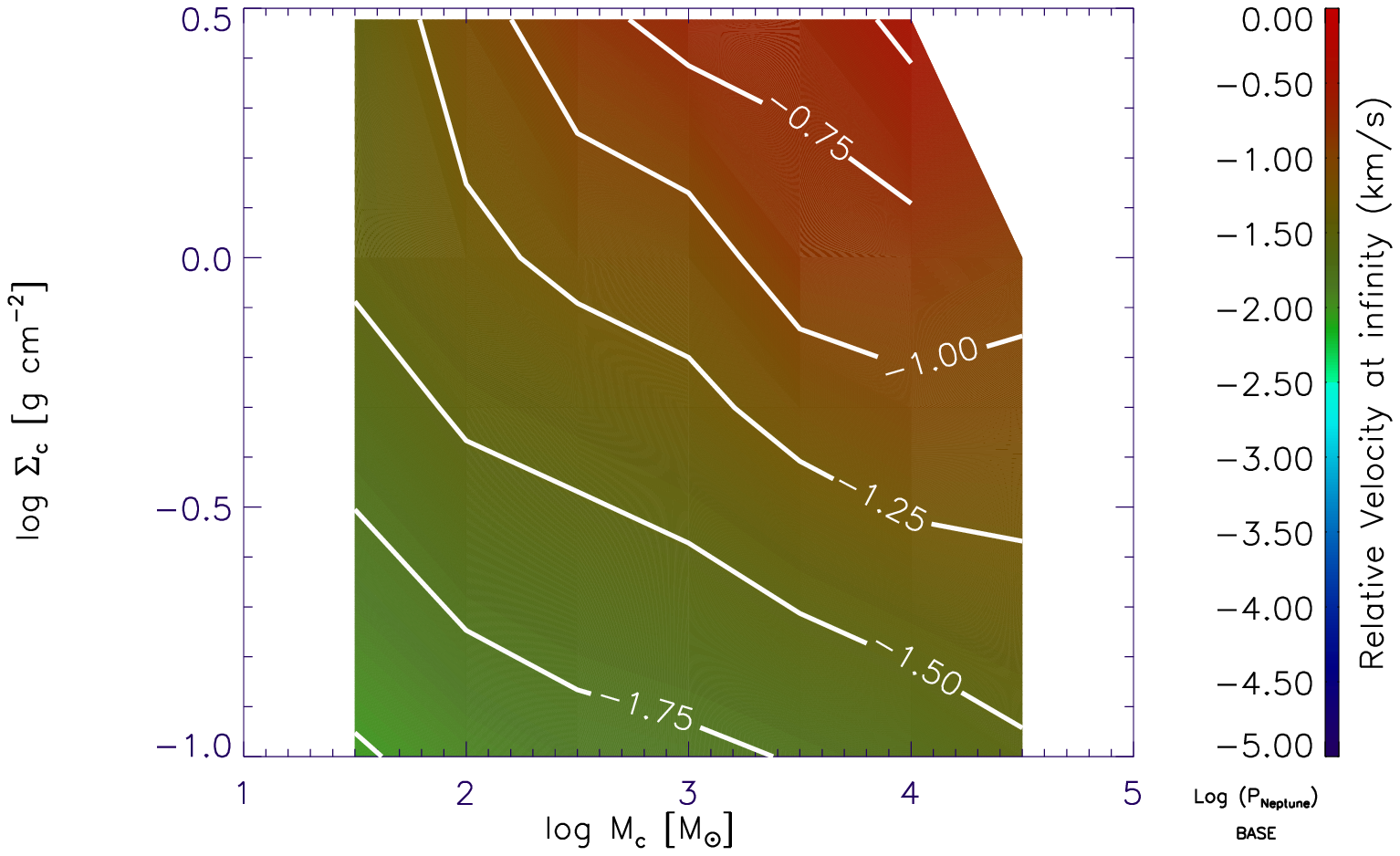}
\plottwo{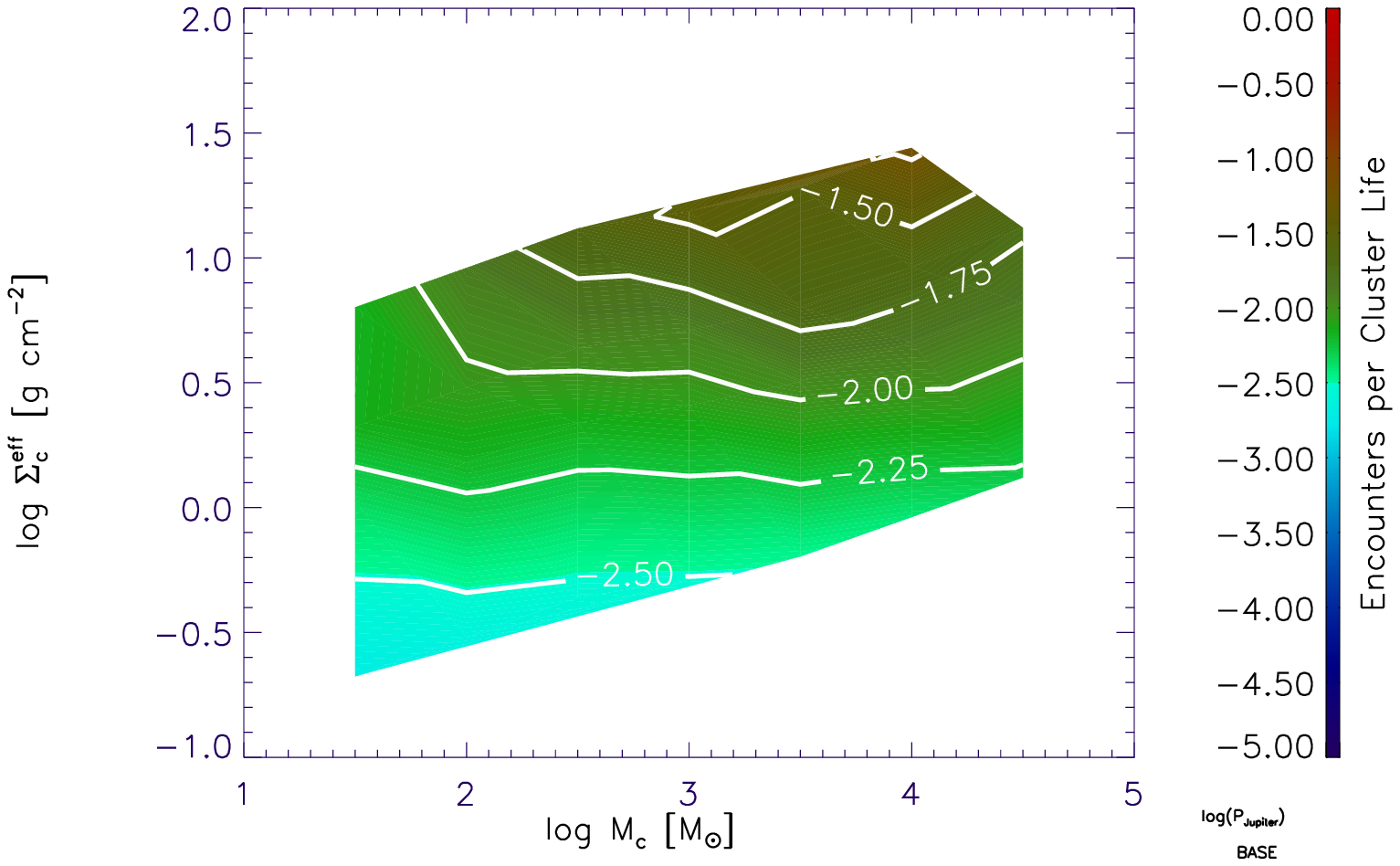}{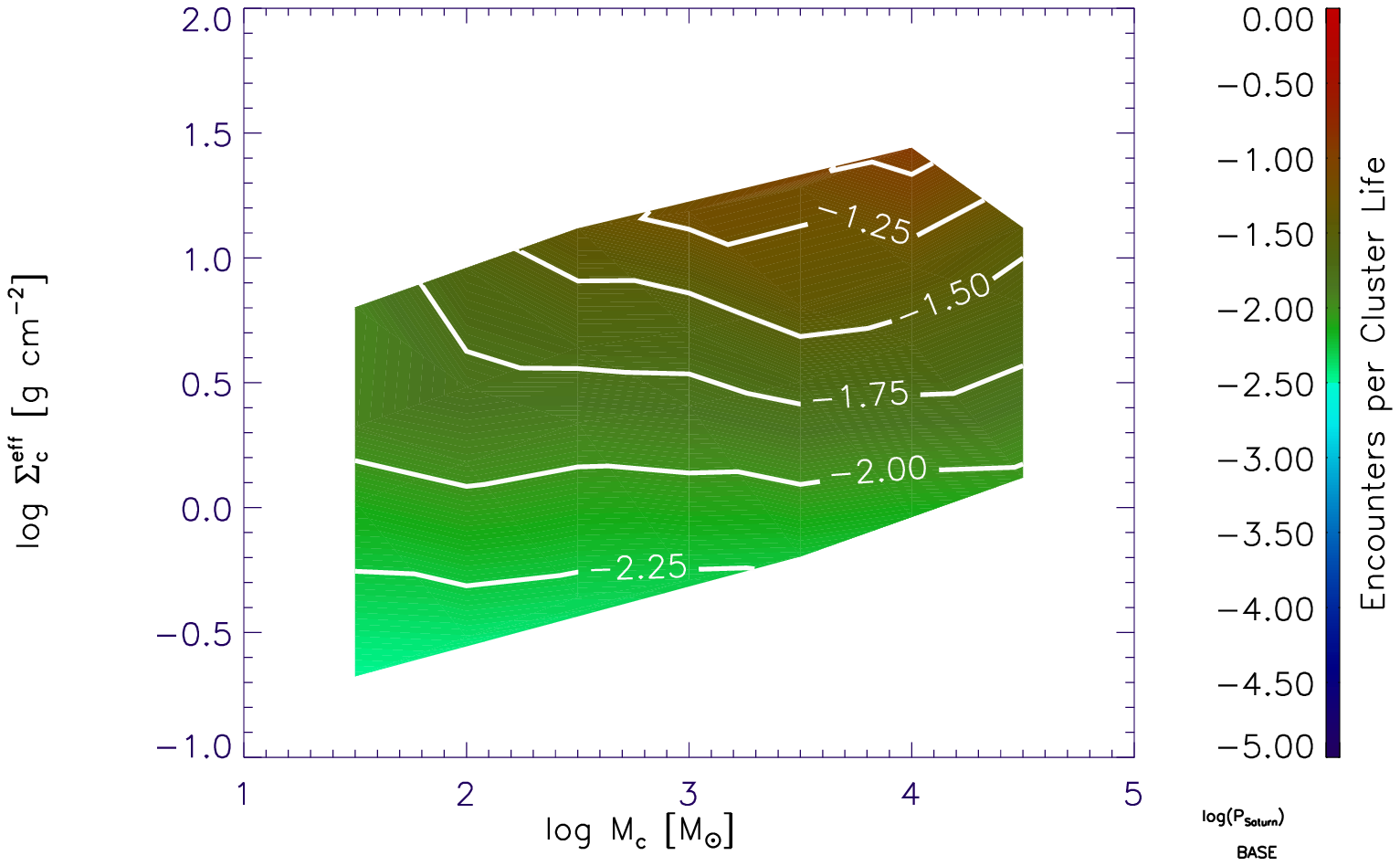}
\plottwo{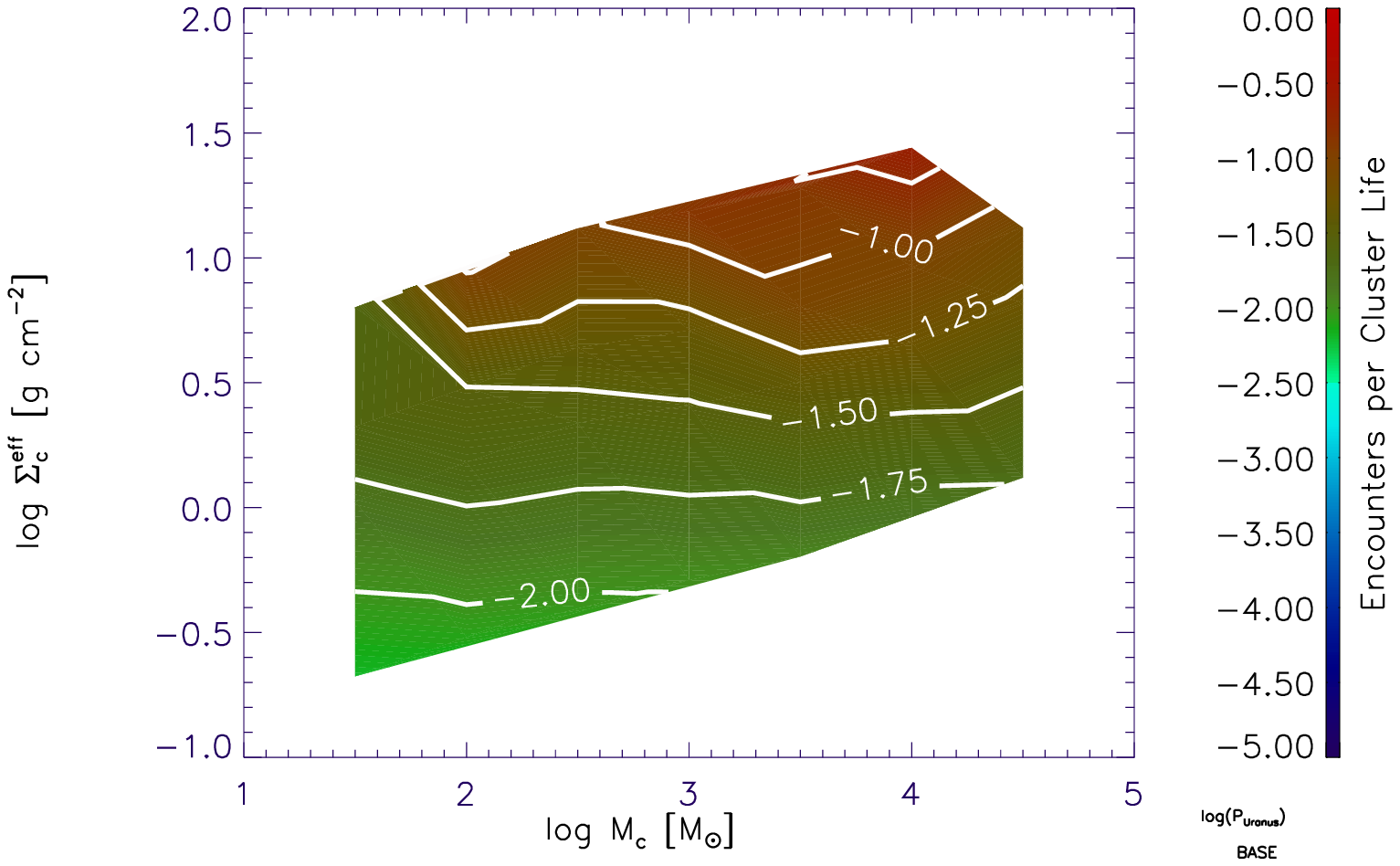}{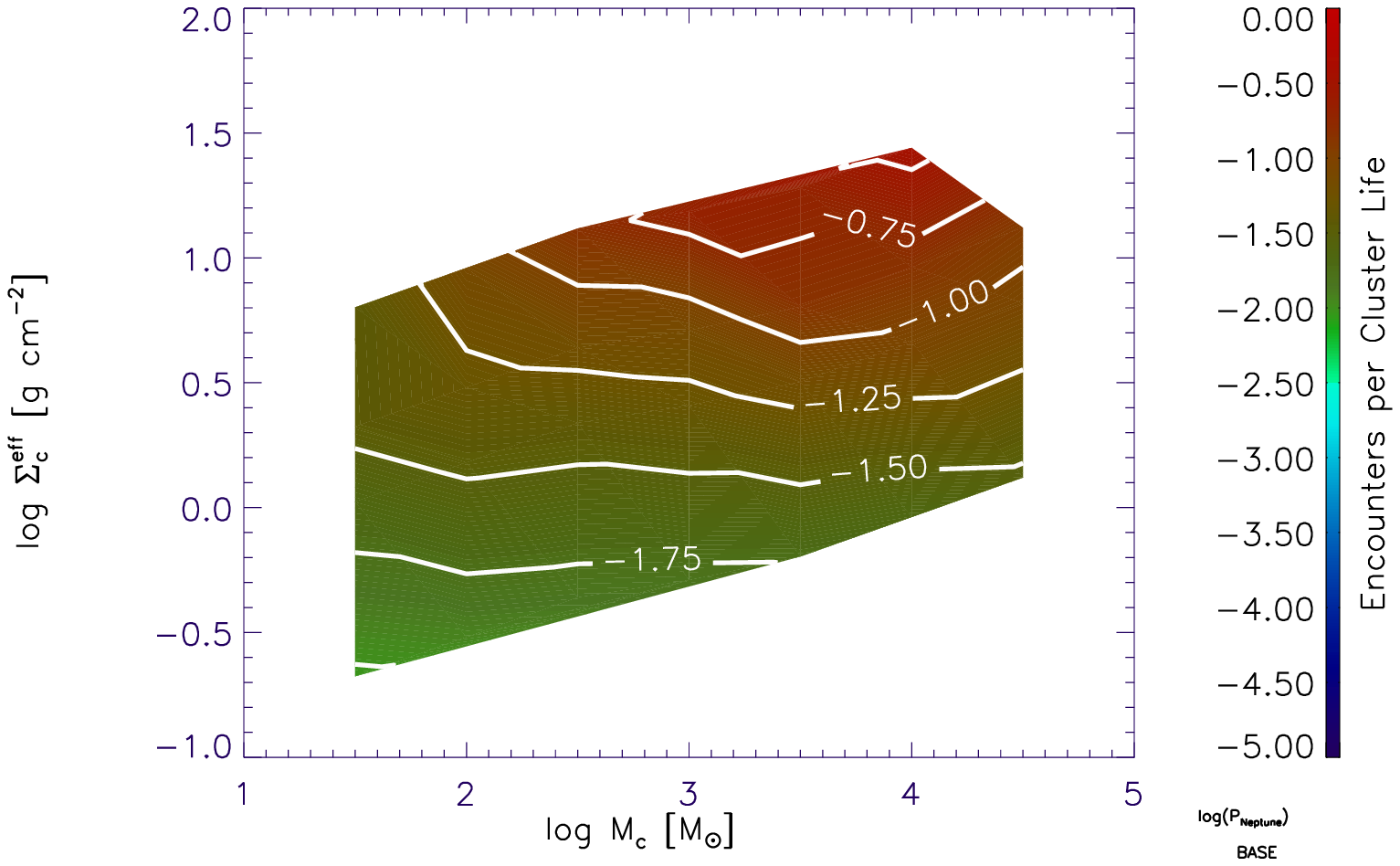}
\caption{The log of the probability that a Jovian planet is excited to an eccentricity $e > 0.1$ as a function of $\Sigma_c$ (top 4) and  $\Sigma_{c,\rm eff}$ (bottom 4) and $M_c$, for model $Q=0.75$, $D=2.2$. Within each set of four we have Jupiter (top left), Saturn (top right), Uranus (bottom left) and Neptune (bottom right).
\label{fig:Q075D22pcombeff}}
\end{figure}

Figure \ref{fig:Q075D22pcombeff} shows the probability of exciting a Jovian planet to eccentricity $e>0.1$ in a cluster with $Q = 0.75$, $D = 2.2$, as a function of $M_c$ and $\Sigma_c$. Overall our conclusions are consistent with those of \citet{Dukes}: even for clusters of quite high masses and surface densities (even up to $\Sigma_{c,\rm eff} = 20$ g cm$^{-2}$), it is extremely unlikely that a Sun-like star would have an encounter with another star close enough to significantly perturb the orbit of a planet like Neptune. However, we also find that the probability of excitation is independent of $M_c$ at fixed $\Sigma_{c,\rm eff}$, but that it increases with $M_c$ at fixed $\Sigma_c$. This is the opposite of the trend found by \citet{Dukes}, and the difference is easy to understand: for a relaxed cluster, as assumed by \citeauthor{Dukes}, the number of encounters is independent of $M_c$ at fixed $\Sigma_c$, while the typical encounter velocity increases with mass as $M_c^{0.25}$, reducing the effective cross-section per encounter. For unrelaxed clusters, on the other hand, we find that the number of encounters increases with $M_c$ and fixed $\Sigma_c$, and that the typical encounter velocity increases with $M_c$ more slowly than is the case for a relaxed system. These two changes reverse the trend predicted by \citeauthor{Dukes}. However, while the direction of trend with mass is reversed, the trend remains rather weak. As a result, the qualitative conclusion that encounters in dispersing clusters should have no significant impact on the Solar system, even if those clusters are quite massive, continues to hold.

\begin{figure}
\plotone{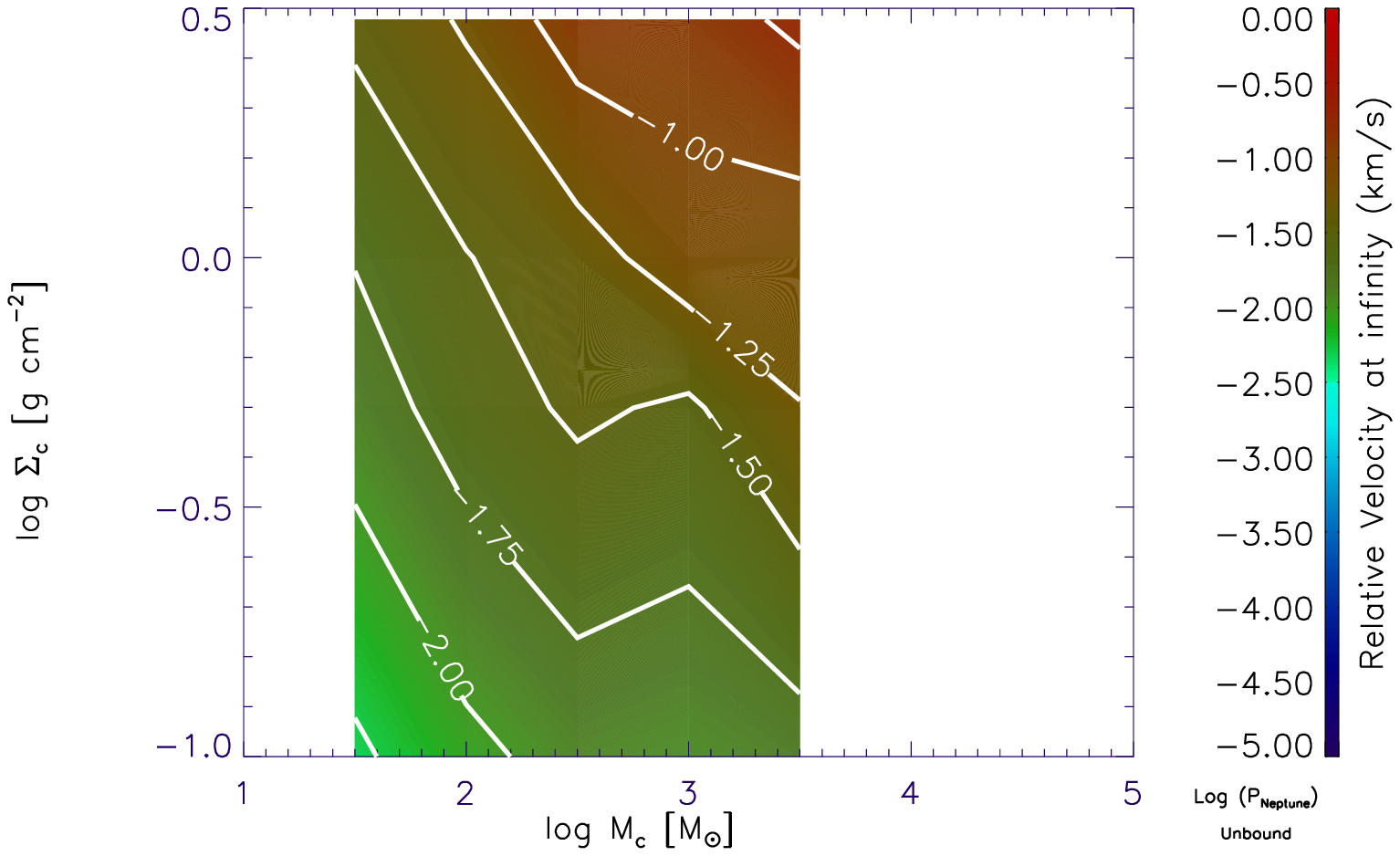}
\plotone{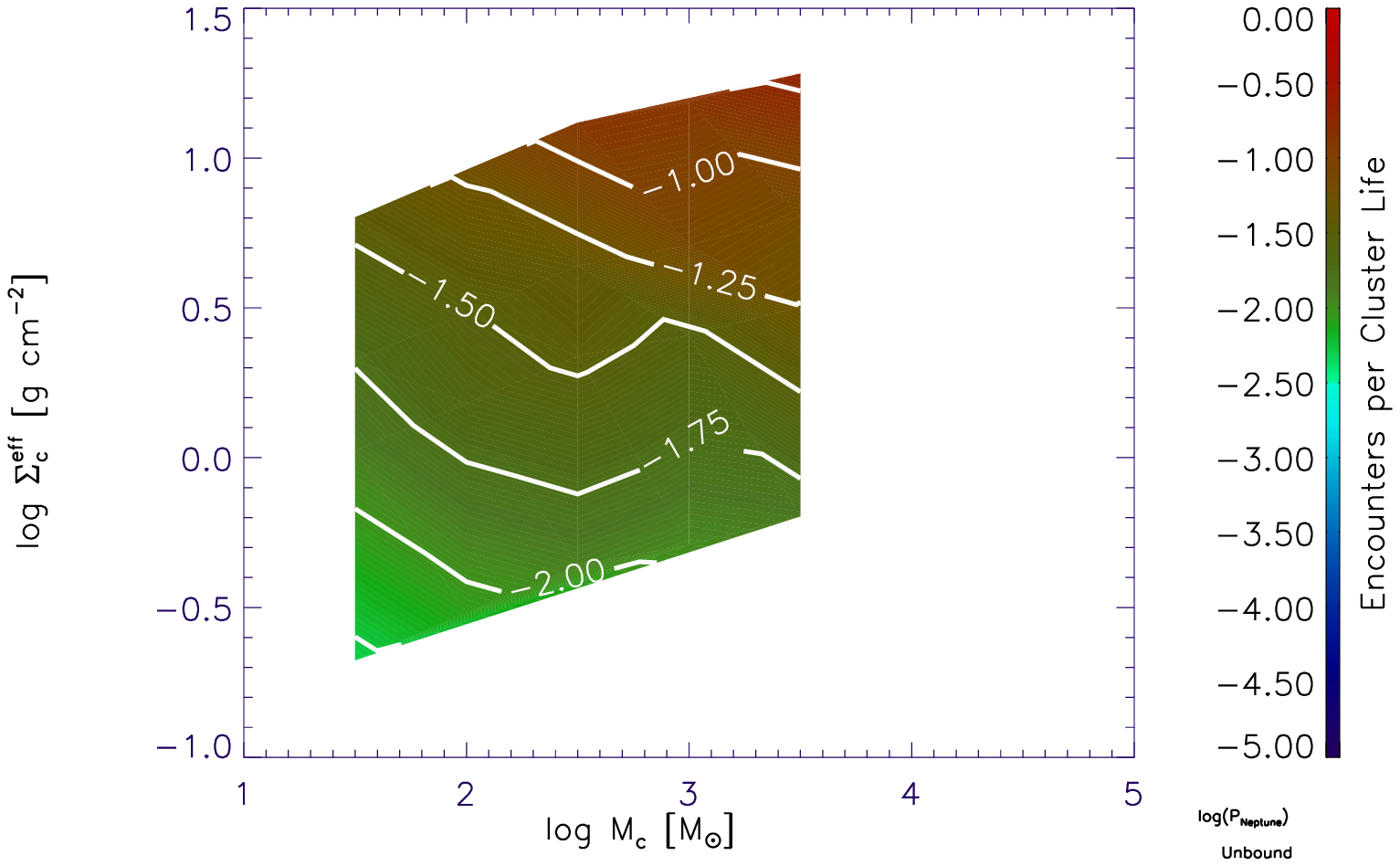}
\caption{The log of the probability of exciting Neptune to an eccentricity $e > 0.1$ as a function of $\Sigma_c$ (top) and  $\Sigma_{c,\rm eff}$ (bottom) and $M_c$, for model $Q=1.25$, $D=2.2$. Notice that the probabilities are lower than those in the unbound case. The other Jovian planets have similar plots to these. 
\label{fig:PNEPunbound}}
\end{figure}

\begin{figure}
\plotone{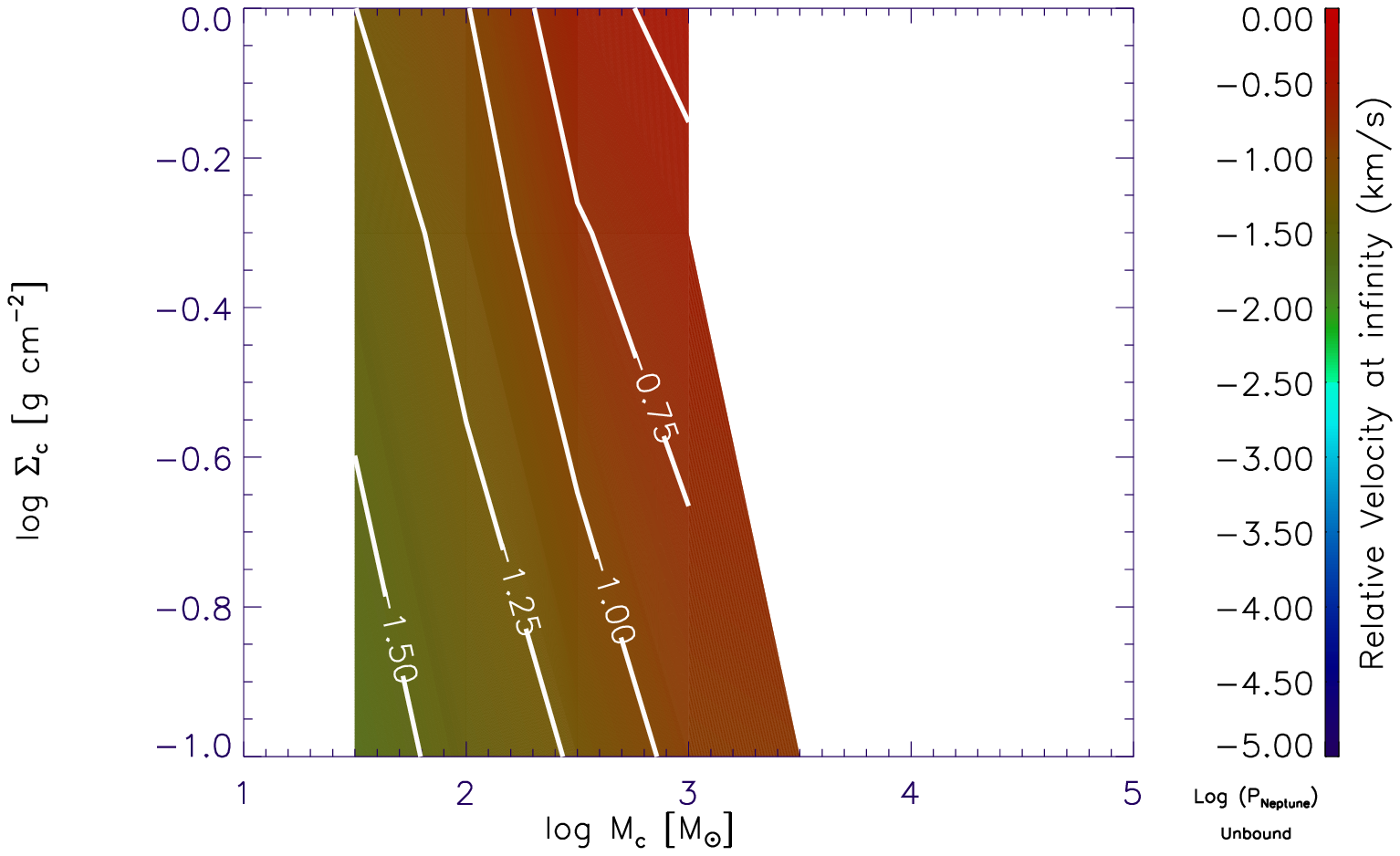}
\plotone{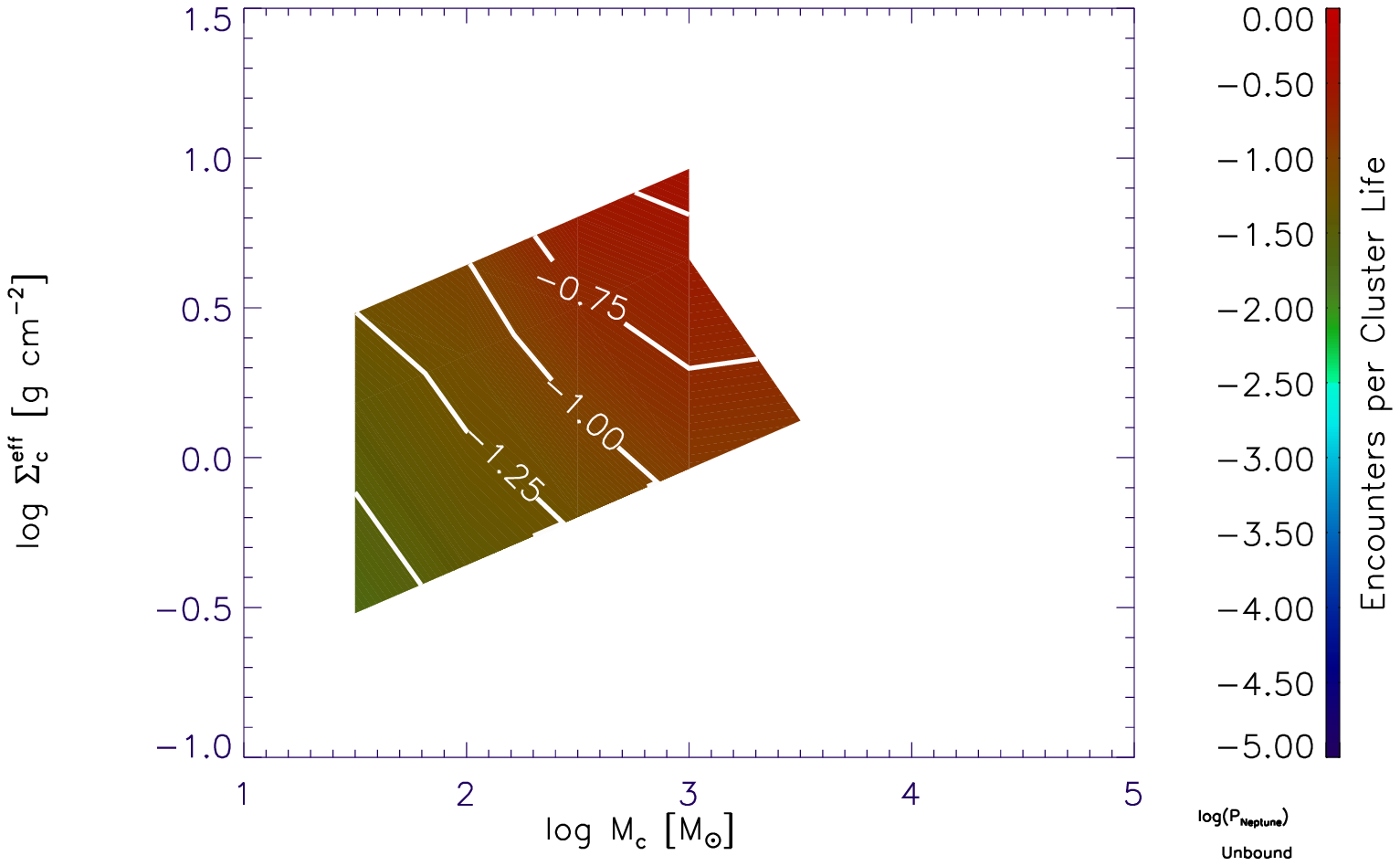}
\caption{Same as figure \ref{fig:PNEPunbound} but in the case of high substructure ($D=1.6$, $Q=0.75$). The other Jovian planets have similar plots, but with a lower excitation probability.
\label{fig:PNEPsubstru}}
\end{figure}

In the unbound case we find similar results to the base case. The combined results of a significant drop in the number of encounters, with a slight decrease in the median relative velocity leads to a slightly lower excitation probability. Figure~\ref{fig:PNEPunbound} shows the probability of exciting Neptune's orbit. Qualitatively the results are similar to those in figure~\ref{fig:Q075D22pcombeff}. The same is true for the high substructure case, model Q0.75D1.6, shown in Figure \ref{fig:PNEPsubstru}. Compared to the base case, the effect of increased number of encounters is much stronger than the increase in relative velocities, leading us to higher excitation probabilities. Again the shape is similar, but slightly more extreme than the base case. The overall probability remains low at low masses, but seems likely to become significant at masses above $\sim 10^4$ $M_\odot$, provided that nominal (as opposed to effective) surface densities remain roughly constant. Whether such highly substructured, massive clusters exist in nature is uncertain.

\begin{figure}
\plotone{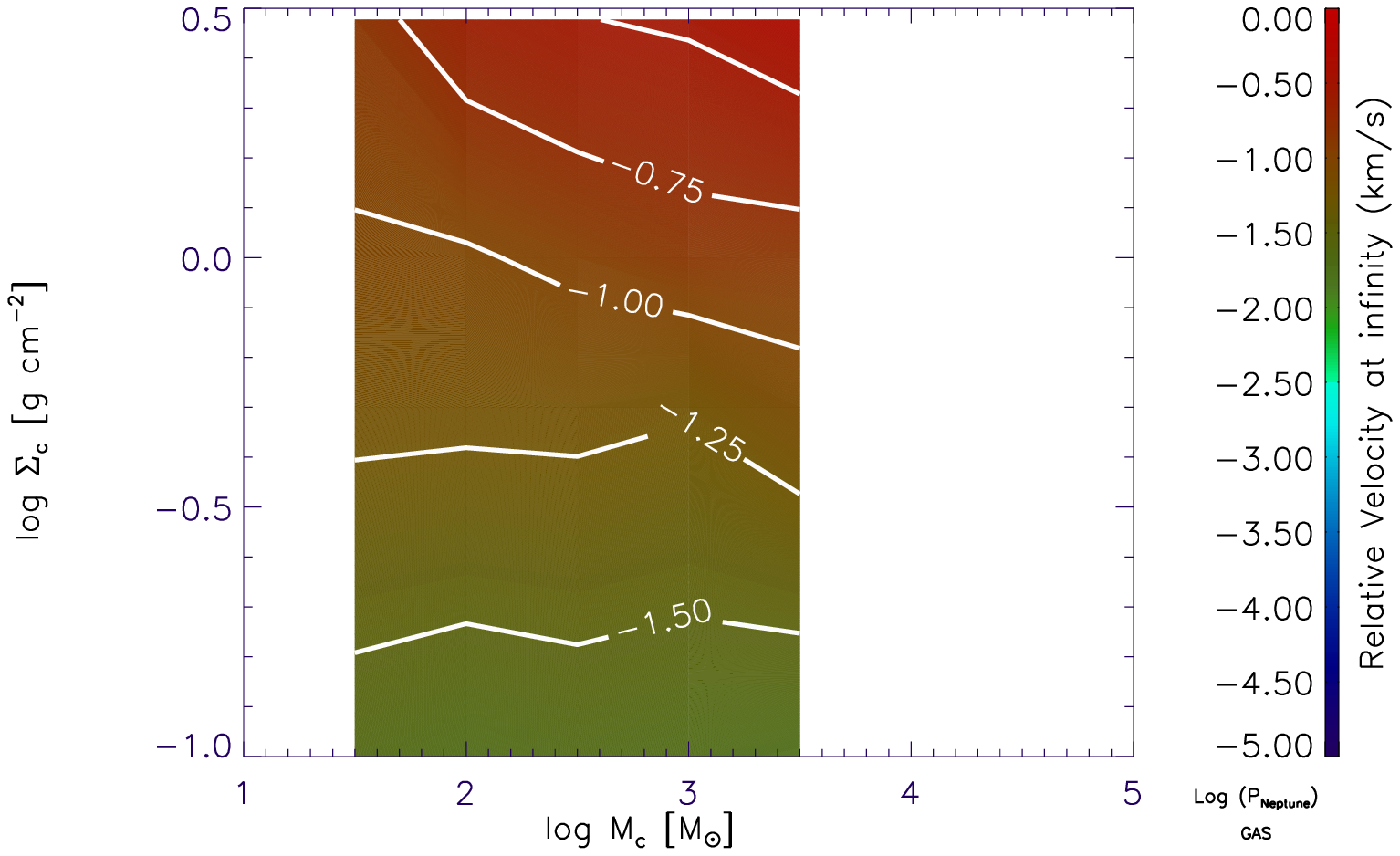}
\plotone{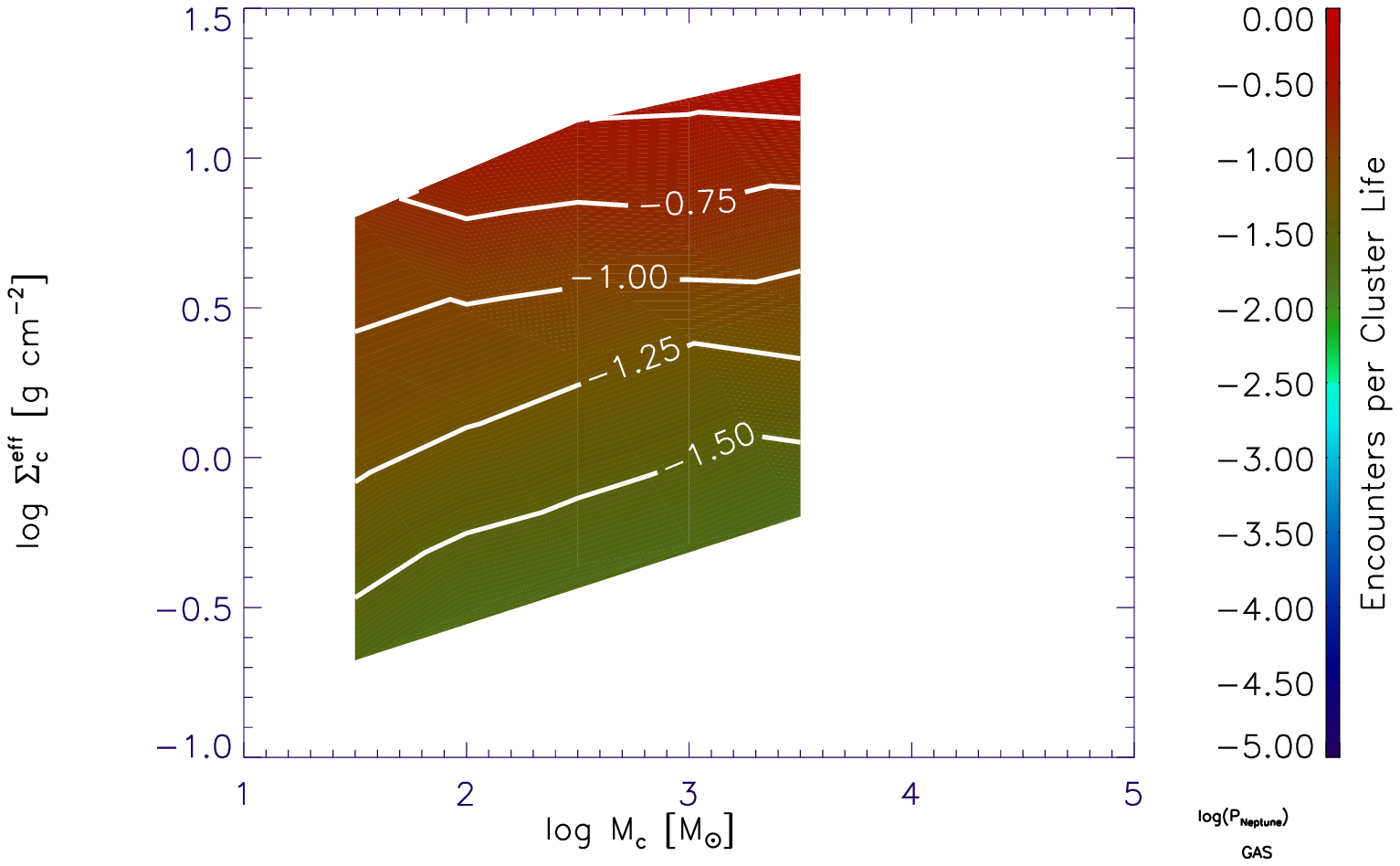}
\caption{Same as figure~\ref{fig:PNEPunbound} but for the case where we include an external (gas) potential for 4 crossing times, and then allow the cluster to disperse.
\label{fig:PNEPgas}}
\end{figure}

The most interesting effects are seen in the Gas case. Due to the destruction of substructure because of the subvirial nature of the cluster, this case approaches that of \citet{Dukes}. Figure~\ref{fig:PNEPgas} shows the probability of exciting Neptune's orbit in such a cluster. The increase in probability of disruptive events with cluster mass is very weak in this case, and when we consider the effective surface density we are able to reproduce the trend of \citeauthor{Dukes}. Namely, at constant (effective) surface density, we find that the probability of a disruptive event actually decreases with the cluster mass. This implies that there is no dynamical limit on the number of stars in the stellar birth cluster. The overall disruption probabilities are slightly higher than those of the base case, but this is to be expected since the cluster undergoes an initial period of highly elevated encounter rate before relaxing and then dispersing.

We can also estimate the errors on these probabilities as follows. The error on the probability of any event (e.g.~excitation of a Jovian planet) is given by
\begin{equation}
\label{perr}
\delta P_i = \exp (- \Lambda_i) \delta \Lambda_i .
\end{equation}
With a little algebra one may show that
\begin{equation}
\label{lamerr}
(\delta \Lambda_i)^2 = \Lambda_i^2 \left[ \left(\frac{\delta N_{enc}}{N_{enc}}\right)^2 + \left(\frac{\delta I_1}{I_1}\right)^2 + \left(\frac{\delta I_2}{I_2}\right)^2\right] ,
\end{equation}
where $I_1$ and $I_2$ are the integrals in the numerator and denominator of equation \eqref{pint}, respectively. 

The errors on $I_1$ and $I_2$ may be obtained by differentiation of the Riemann sum approximating the integral:
\begin{equation}
\label{I1err}
(\delta I_1)^2 = \sum_i v_i^2 \sigma_i^2 p_i^2 \left[\frac{(\delta \sigma_i)^2}{\sigma_i^2} + \frac{1}{p_i}\right](\Delta v_i)^2.
\end{equation}
For the sake of compactness of the expression we omit the $\infty$ subscript on the velocity. $I_2$ is given by a similar expression, except that the cross section is absent. Given the errors on $N_{\rm enc}$ (see Section \ref{sec:error}), $I_1$, and $I_2$, we can compute the overall errors on our probabilities from \eqref{perr}.

To keep this analysis as simple as possible we assume that the relative errors on the integrals are very small, so that the relative error on the number of encounters dominates. Then we have 
\begin{equation}
\delta \Lambda _ i = \Lambda_i \frac{\delta N_{\rm enc}}{N_{\rm enc}} .
\end{equation}
Typically $\Lambda_i \ll 1$ (all cases except one have $\Lambda_{\rm Neptune} < 10^{-2}$, and all of the other Jovian planets have smaller values) and in this case we can expand equation \eqref{perr} to first order in $\Lambda_i$ to obtain
\begin{equation}
\delta P_i \approx \Lambda_i \frac{\delta N_{\rm enc}}{N_{\rm enc}} = \Lambda_i \sigma_r.
\end{equation}
Since typical values of $\Lambda_i$ are $O(10^{-2})$ or smaller, and typical values or $\sigma_r$ are also $O(10^{-1})$ or smaller, this means that the absolute error on the percentages calculated through this method are less than $0.1\%$, and the relative error, of order $\sigma_r$, is at most $\sim 10\%$. Thus we can be fairly confident in our major results. 

Finally, there a few additional sources of error that we mention here, but do not quantify explicitly. The IMF used in \citet{Dukes} is slightly different than ours, and yields a mean stellar mass of approximately $0.2 M_{\odot}$ instead of $0.59M_\odot$. This implies that we would expect the value of the probabilities to be slightly higher, although the shape of the constant probability curves would not change (figure \ref{fig:Q075D22pcombeff}). Similarly, the fact that there is slight deviation from the distribution $P(b) \propto b$, especially at high $\Sigma_c$, is a source of error. At low surface density, however, this error should be minimal. To obtain an entirely correct result for $\Lambda_i$ one should repeat the calculations of \citet{Dukes} using the same IMF and distribution of impact parameters as found in these simulations.

\subsection{Free-Floating Planets}

Ever since their discovery by \citet{2011Natur.473..349S}, there has been considerable debate about the origin of planetary mass objects that are either completely unbound or very distant from their parent stars. Models for the origin of these objects include planet-planet scattering leading to ejection in a young planetary system \citep[e.g.][]{2011ApJ...742...72N, 2012ApJ...754...57B}, escape caused by mass loss during late stages of stellar evolution \citep[e.g.][]{2011MNRAS.417.2104V}, direct formation from the interstellar medium \citep{2012MNRAS.423.1856S}, and dynamical stripping of planets from stars in clusters \citep[e.g.][]{2012ApJ...754...57B, 2012MNRAS.421L.117V, 2012arXiv1209.2136O}. While a full discussion of this topic is beyond the scope of this paper, we do note that our simulations provide new insight into the latter mechanism.

Our results from the previous section show that significant orbital disturbance for a planet at 30 AU, the distance of Neptune, is likely to be a relatively rare event. Only for clusters whose masses exceed $\sim 10^4$ $M_\odot$ \textit{and} are very highly substructured ($D=1.6$) do orbital excitations become common. Complete ejection will be even rarer. This suggests that, unless the planet formation process is capable of producing large numbers of planetary mass objects at radii significantly beyond 30 AU by internal mechanisms (e.g.~planet-planet scattering), cluster stripping cannot be a significant source of free-floating planets. A more precise and quantitative version of this statement may be obtained by combining the orbital element distributions we have obtained here with population synthesis models for star clusters and planets.

Finally, it is interesting to compare the results of our simulations with those of \citet{2012MNRAS.419.2448P}. They find large orbital excitations only in their $Q=0.3$ case, which includes no gas potential term. This will lead to a tightly bound final cluster. This fact together with their long simulation times (10 Myr, or $\sim 30$ crossing times) means that it is not surprising that there are a large fraction of planets becoming unbound. In addition, they consider orbital excitations of planets from stars of any mass, and by number the majority of stars are smaller than the $0.8-1.2 M_{\odot}$ mass range that we consider. The binding energy of planets in such systems will be smaller than is typical of the planetary systems we consider. 

The closest match in parameters between our simulations and those of \citeauthor{2012MNRAS.419.2448P} is between our $Q=0.75, D=2.2$ and their $Q=0.7, D=2.0$ runs. Their figure 8 corresponds roughly to clusters with $M_c = 10^2 - 10^3 M_{\odot}$ and $\Sigma_c=0.1$ g cm$^{-2}$. They find that $\sim 10\%$ of planets at $30$ AU are stripped from their parent star in this case, whereas we find typically that $\sim 2-3\%$ of orbits are excited in such a case. These differences are most likely due to the issue of binding energies mentioned above. For the IMF used by \citeauthor{2012MNRAS.419.2448P}, and their assumption that planets are equally likely to occur around stars of any mass, only $\sim 20\%$ of planets orbit stars of mass $>1$ $M_{\odot}$, while $\sim 40\%$ are born around stars with mass $<0.5$ $M_{\odot}$. Thus, it is not surprising that they find a higher rate of orbital perturbation and stripping. Which model will more accurately predict numbers of free-floating planets is unclear, due to the lack of reliable estimates of planet fractions around stars with mass $\ll 1 M_{\odot}$.  

\section{Summary and Conclusions}
\label{sec:conclusion}

We have conducted a series of simulations of the evolution of dispersing young star clusters that possess a high degree of substructure. These simulations cover a wide range of masses ($30 - 30,000$ $M_\odot$), surface densities ($\Sigma_c = 0.1 - 3.0$ g cm$^{-2}$), dynamical states (supervirial but bound, unbound, and subvirial but then unbound by gas expulsion), and degrees of substructure (fractal dimension $D=1.6$ and $2.2$). These parameters are chosen to reproduce the range of properties for young star clusters suggested by both observations and star formation simulations. We provide tabulated distributions of number of encounters, impact parameters, and relative velocities as a function of these properties. These may be used as inputs in population synthesis models of planet formation.

Our calculations produce a number of interesting conclusions. First, during the $\sim 1$ crossing time it takes for the initial substructure to be erased by dynamical interactions, the number of encounters is significantly elevated compared to what one would expect for a relaxed system. The amount of elevation depends the amount of substructure and other cluster properties. Regardless of its strength, though, the enhancement is transient. Either the substructure dissolves (if the cluster is not confined or mildly bound), or the cluster disperses (if it is strongly unbound). Thus for moderate degrees of substructure the overall enhancement in the number of encounters is only a factor of a few for clusters that do remain bound for some period before gas dispersal. Only if the gas has extremely strong substructure (fractal dimension $D=1.6$) is the enhancement larger.

Second, early in the evolution of a substructured cluster, before the cluster has dispersed or the substructure has been erased, the distribution of encounter impact parameters is not far off from the expectation for a relaxed cluster, but the distribution of velocities is significantly non-Maxwellian. Because this early phase contributes an appreciable fraction of all encounters even in clusters that remain bound for four crossing times, the overall distribution of encounter velocities is non-Maxwellian even in such clusters. Compared to a Maxwellian, our clusters show both a sharper peak at moderate encounter velocities, and a longer tail extending to higher velocities. The overall median velocity increases with cluster mass more slowly than one would expect in a relaxed cluster, and scales with the virial velocity ratio. Clusters with larger virial ratios, and thus larger velocity dispersions, actually tend to produce lower median encounter velocities because they are less effective at dissolving the velocity substructure and randomizing stellar relative velocities.

Third, even with the enhanced encounter rates that we find, we conclude that planetary systems or protoplanetary disks around stars in dissolving clusters are unlikely to experience significant dynamical perturbations from other stars in the cluster, at least for planets that are within tens of AU of their parent star. Such planets are simply too tightly bound and have cross sections that are too small for many of them to be disturbed. This remains true even in our most highly substructured cases, which produce the largest number of encounters, up to cluster masses of $10^{3.5}$ $M_{\odot}$. This means that there is no dynamical constraint on the size of the Sun's parent cluster, and that cluster stripping is unlikely to be an important contributor to the population of free-floating planets in the Milky Way.

\acknowledgements We thank D.~Dukes for help running simulations and analyzing the data, and S.~Aarseth for assistance with NBODY6. MRK acknowledges support from the Alfred P.~Sloan Foundation,  the NSF through grant CAREER-0955300, and NASA through Astrophysics Theory and Fundamental Physics Grant NNX09AK31G, and a Chandra Space Telescope Grant.

\begin{appendix}

\section{Derivation of an Effective Surface Density for Fractal Star Clusters}

We can define an effective surface density $\Sigma_{c,\rm eff}$ by considering the process by which the fractal cluster is generated. Let the cube out of which the fractal is built have a volume given by $V_0$, with a characteristic radius $r_c$. The process of building the fractal can be thought of as removing chunks of the volume from this initial value. Since the first generation particles are always parents, the volume loss starts at the second generation. The volume lost after the second generation is, on average, $V_{\rm loss,2}=V_0 (1 - 2^{D-3})$. The volume remaining after the second generation is then $V_{\rm remain,2}= V_0 - V_{\rm loss,2} = V_0 2^{D-3}$. Similarly, the volume lost after the third generation is $V_{\rm loss,3} = V_{\rm remain,2} (1 - 2^{D-3})$, the remaining volume is $V_{\rm remain,3}=V_0 (2^{D-3})^2$, and so forth. The volume remaining after $g$ generations is simply
\begin{equation}
V_{{\rm remain},g} = V_0 2^{(D-3)(g-1)}.
\end{equation}
We can thus define an effective radius by
\begin{equation}
r_{c,\rm eff} = r_c 2^{\frac{1}{3}(D-3)(g-1)}.
\end{equation}
Here we have implicitly assumed that we have taken enough generations of the fractal that when we make the distribution of positions spherical, that the volume loss remains the same. By using dimensional scaling arguments we see that the effective surface density is then
\begin{equation}
\label{sigeff}
\Sigma_{c,\rm eff}(D,g) = \Sigma_c^0 2^{-\frac{2}{3}(D-3)(g-1)}.
\end{equation}
where $\Sigma_c^0$ is the surface density of a cluster with constant density of stars and radius $r_c$. The effective initial surface density therefore depends on how we choose the number of generations for our fractal, which in turn is determined by the mass of the cluster, since the number of generations required is determined by the condition that there be enough potential sites to accommodate the number of stars in the cluster. We approximate this condition by
\begin{equation}
g(N) \approx \frac{\ln(2N)}{\ln(8)} + 1 + s_2(D), 
\end{equation}
where $N=M_c/\bar{m}$ is the number of stars in the cluster and $s_2(D)$ is $1$ if $D<2$ and $0$ otherwise. This choice comes from the fact that at $D=2$, $P_{\rm parent}=0.5$. The average number of generations required to generate the fractal as a function of the mass is displayed in table~\ref{gen}. We obtained this result by averaging over a random sample of 100 different fractals for each mass bin.


\begin{deluxetable}{crrrrrrr}
\tablecaption{
\label{gen}
Number of Generations Versus Cluster Mass
}
\tablehead{
\colhead{} &
\multicolumn{7}{c}{$\log (M_c/M_\odot)$ } \\
\colhead{D} &
\colhead{1.5} &
\colhead{2.0} &
\colhead{2.5} &
\colhead{3.0} &
\colhead{3.5} &
\colhead{4.0} &
\colhead{4.5}
}
\startdata
2.2 & 3.02 & 4.00 & 4.99 & 5.42 & 6.01 & 7.00 & 7.98 \\
1.6 & 3.99 & 4.98 & 5.99 & 7.01 & 8.00 & -- & --
\enddata
\tablecomments{The average number of generations required to generate a cluster of mass $M_c$, for $D=2.2$ and $D=1.6$.}
\end{deluxetable}

\end{appendix}
\bibliographystyle{apj}
\bibliography{thesisbib}
\end{document}